

\input harvmac.tex

%

\def\bar{\overline}
\def\hat{\widehat}
\def\*{\star}
\def\({\left(}		\def\BL{\Bigr(}
\def\){\right)}		\def\BR{\Bigr)}
\def\[{\left[}		\def\BBL{\Bigr[}
\def\]{\right]}		\def\BBR{\Bigr]}

%
%
\def\frac#1#2{{#1 \over #2}}		\def\x{ \otimes }
\def\inv#1{{1 \over #1}}
\def\half{{1 \over 2}}
\def\d{\partial}
\def\der#1{{\partial \over \partial #1}}

\def\vev#1{\langle #1 \rangle}
\def\ket#1{ | #1 \rangle}
\def\bra#1{ \langle #1 |}

\def\2pi{\hbox{$2\pi i$}}

\def\dsl{\raise.15ex\hbox{/}\kern-.57em\partial}
\def\Dsl{\,\raise.15ex\hbox{/}\mkern-.13.5mu D}
%
%
\def\th{\theta}		\def\Th{\Theta}
\def\ga{\gamma}		\def\Ga{\Gamma}
\def\be{\beta}
\def\al{\alpha}
\def\ep{\epsilon}
\def\la{\lambda}	\def\La{\Lambda}
\def\de{\delta}		\def\De{\Delta}
		\def\Om{\Omega}
\def\sig{\sigma}	

%
%
		\def\CC{{\cal C}}
\def\CD{{\cal D}}		\def\CF{{\cal F}}
\def\CG{{\cal G}}		\def\CI{{\cal J}}
\def\CJ{{\cal J}}		
		\def\CO{{\cal O}}
\def\CP{{\cal P}}		\def\CR{{\cal R}}
	\def\CT{{\cal T}}

%
%
\font\numbers=cmss12
\font\upright=cmu10 scaled\magstep1
\def\stroke{\vrule height8pt width0.4pt depth-0.1pt}
\def\topfleck{\vrule height8pt width0.5pt depth-5.9pt}
\def\botfleck{\vrule height2pt width0.5pt depth0.1pt}
\def\Zmath{\vcenter{\hbox{\numbers\rlap{\rlap{Z}\kern 0.8pt\topfleck}\kern
2.2pt
                   \rlap Z\kern 6pt\botfleck\kern 1pt}}}
\def\Qmath{\vcenter{\hbox{\upright\rlap{\rlap{Q}\kern
                   3.8pt\stroke}\phantom{Q}}}}
\def\Nmath{\vcenter{\hbox{\upright\rlap{I}\kern 1.7pt N}}}
\def\Cmath{\vcenter{\hbox{\upright\rlap{\rlap{C}\kern
                   3.8pt\stroke}\phantom{C}}}}
\def\Rmath{\vcenter{\hbox{\upright\rlap{I}\kern 1.7pt R}}}
\def\Z{\ifmmode\Zmath\else$\Zmath$\fi}
\def\Q{\ifmmode\Qmath\else$\Qmath$\fi}
\def\N{\ifmmode\Nmath\else$\Nmath$\fi}
\def\C{\ifmmode\Cmath\else$\Cmath$\fi}
\def\R{\ifmmode\Rmath\else$\Rmath$\fi}
\def\oti{\x}
\def\g{ {\cal G} }
\def\Ad{ {\rm Ad} }

\Title{SPhT-91-124 }{\vbox{
\centerline{Quantum Symmetries in 2D Massive
Field Theories.}}}
\bigskip
\bigskip

\Date{Cargese-91}

\rightline{SPhT-91-124 }
\bigskip \bigskip
\vbox{
\centerline{{\bf QUANTUM SYMMETRIES }}
\centerline{{\bf IN 2D MASSIVE FIELD THEORIES.}}}
 {}\footnote.{Series of lectures given at the 91 Cargese school,
{\it New symmetry principles in quantum field theory}.}
\bigskip \bigskip
\centerline{Denis BERNARD}
\bigskip
\centerline{Service de Physique Th\'eorique de Saclay
\footnote*{Laboratoire de la Direction des sciences de la mati\`ere
du Commissariat \`a l'\'energie atomique.}}
\centerline{F-91191 Gif-sur-Yvette, France.}
\bigskip \bigskip
\bigskip \bigskip
\bigskip \bigskip
\noindent We review various aspects of (infinite) quantum
group symmetries in 2D massive  quantum field theories.
We discuss how these symmetries can be used
to exactly solve the integrable models. A possible way for
generalizing to three dimensions is shortly described.
\vfill \eject
 ~\bigskip
\noindent {\bf Content:}\par
\bigskip
\noindent
1 - Quantum Symmetries in 2D Lattice Field Theories.\par
1a- {\it Quantum symmetries and conserved currents.}\par
1b- {\it Fields multiplets.}\par
1c- {\it Examples.}\par

\noindent 2 -
Classical Origin of Quantum Symmetries: Dressing Transformations.\par
2a- {\it What are the dressing transformations.}\par
2b- {\it Few of their properties.}\par

\noindent
3 - An Example of Dressing Transformations: Current Algebras.\par
3a- {\it The equations of motions.}\par
3b- {\it Dressing transformations and the Riemann-Hilbert problem.}\par
3c- {\it Non-local conserved currents.}\par

\noindent
4 - Quantification: Yangians in Massive Current Algebras.\par
4a- {\it OPE's in massive quantum current algebras.}\par
4b- {\it The quantum non-local currents.}\par
4c- {\it The non-local conserved charges and their algebra.}\par
4d- {\it Action on the asymptotic states and the S-matrices.}\par
4e- {\it Action on the fields and the field multiplets.}\par

\noindent
5 - Conclusions.\par
{\it 3D generalizations ??}

\vfill \eject

\noindent {\bf INTRODUCTION }

Symmetries of the S-matrices of 4D quantum field theories
are subject to the severe constraints of the Coleman-Mandula theorem
\ref\rCM{S. Coleman and J. Mandula, Phys. Rev. 159 (1967) 1251}.
In general, these possible symmetries do not
allow for non-perturbative solutions of the theories.
In two dimensions, this theorem breaks down and there is room
for richer symmetries. The aim of the
lecture is to describe new symmetries of 2D  massive QFT,
known as quantum group symmetries.  This paper is mainly a review of
published papers completed by few remarks and comments.

There are at least two motivations for studying quantum symmetries in
2D quantum field theories:\par
\noindent {\bf (i) The study of the possible symmetries of 2D QFT.}
The quantum group symmetries we will
analyse in these lectures are characterized by the fact that,
unlike standard Lie algebra symmetries, they do not act
additively, and by the fact that they are generated by
non-local currents having in general non-integer Lorentz spins.
They thus provide non-Abelian extension of the Lorentz group. \par

\def\hb{ {\hat \beta} }
\def\slq{ {\hat {sl_q(2)} }}
\def\bQ{ {\bar Q} }
\def\zb{{\bar z}}
\def\bh{{ \hat{\beta} }}

\noindent {\bf (ii) An algebraic formulation of 2D QFT
and exact solutions.}
Our (desesperate ?) goal is to formulate an algebraic approach to 2D QFT,
based on their symmetries (local and non-local),
which could offer a way to solve the integrable two-dimensional
quantum field theories from symmetry data, in analogy with
the approach used in conformal field theories
\ref\rBPZ{A. Belavin, A. Polyakov and A. Zamolodchikov,
Nucl. Phys. B241 (1984) 333}.

It is of some interest to compare the approach which as been
used recently in CFT and in 2D integrable models.
a) Rational conformal field theories are invariant under chiral
vertex operator algebras, which could be local, e.g. the Virasoro or affine
algebras, or non-local, e.g. the parafermionic algebras.
The conformal field theories are reformulated as representation
theories of the chiral algebras: Hilbert spaces of the CFT's are direct sums
of representations of the chiral algebras;
conformal primary fields are intertwiners for the
chiral algebras, etc...
The chiral algebras are non-abelian and this is, for a large part,
at the origin of the exact solvability of the CFT's.
Being completely integrable, the CFT's also possess infinitely many
local integrals of motion in involution. However, almost none of these
integrals of motion are actually used to solve the conformal models. (It is
even difficult to express them
in terms the generators of the chiral algebras.)
b) In contrast, a way of studying integrable
models \ref\rInt{See e.g. O. Babelon and C.-M. Viallet, {\it
Integrable models, Yang-Baxter equation and quantum groups},
LPTHE-Paris preprint (1989), to appear (?) as a book; or
L.D. Faddeev and L. Takhtadjan, {\it Hamiltonian methods in theory of
solitons},
Springer Verlag (1987).}
consists in extracting local integrals of motion which are in
involution. These integrals of motion  thus form an
abelian algebra. There existence ensures the integrability
of the theory and the factorization of the S-matrix.
In general, they do not provide enough informations for
solving the theory, e.g. for determining the S-matrix.

Thus, almost none of the techniques used in one of these
fields is used in the other. However,
besides these local integrals of motion, the 2D integrable models
and the conformal models
also possess non-local integrals of motion. These non-local conserved
charges are the generators of non-abelian algebras known as the
quantum symmetry algebras of the models. It is hoped that
these new symmetry algebras will allow us to define a framework
which could be apply simultaneously to the conformal field
theories and to the massive integrable models.

To characterize the massive quantum field theories
uniquely by their symmetry algebras requires :\par
\noindent (I) that the asymptotic particles form multiplets
of the symmetry algebras and that the invariance for the
S-matrix determine it uniquely. As we will see
in the course of this lecture, because quantum group symmetries in 2D
do not commute with the Lorentz group, they provide
algebraic relations on the S-matrices. \par
\noindent (II)
that all the fields of the QFT can be gathered into
field multiplets transforming covariantly under the symmetry
algebras and that the intertwining
properties for the field multiplets determine them uniquely
(in analogy with the minimal assumption in rational conformal field theories).
This amounts to demand that the Ward identities have unique solutions.
As we will describe, the components of the field multiplets,
the descendents and the highest vector fields,
are related through the Ward identities, once more in complete
analogy with conformal field theories.

The symmetry algebras having these requirements
could be called complete symmetry
algebras. For a given model there could be more than one complete
symmetry algebra. The problem of solving
the integrable massive models reduce to the problem of finding
a complete symmetry algebra. As we already said, the local
integrals of motion which are involution do not form (in general)
a complete symmetry algebra. The quantum symmetry we are going to
describe provide in general more informations than these abelian
integrals of motion. To our knownledge, it is not known if they form or not
a complete symmetry algebra. However, it is tempting
to conjecture that the algebra generated by the local integrals of
motion together with the generators of the quantum symmetry form
a complete symmetry algebra.

\bigskip
\noindent {\bf An example:  $\slq$ -symmetry in the Sine-Gordon models.}
The quantum sine-Gordon theory is described by the Euclidean action
\eqn\IIIi{ S\ =\ \inv{4\pi}\int d^2z\ \BBL\d_z\Phi \d_{\zb}\Phi \
+ \la\   :\cos\(\hb \Phi\): \BBR~~.}
The parameter $\hb$ is a coupling constant; it is related to the
conventially normalized coupling by $\hb=\be /\sqrt{4\pi}$.
For $\hb \leq \sqrt{2}$ the action can be renormalized
by normal-ordering the $\cos\( \hb \Phi\)$
interaction and absorbing the infinities into $\la$; the coupling
constant $\bh$ is thereby unrenormalized
\ref\rCol{S. Coleman, Phys. Rev. D11 (1975) 2088}.
The sine-Gordon theory has a well known topological current:
$ J^{\mu}(x,t) = \frac{\hb}{2\pi}\ \ep^{\mu\nu}
        \d_{\nu} \Phi(x,t)$
where $\ep^{\mu\nu}=-\ep^{\nu\mu}$.
The topological charge is:
\eqn\IIIvi{\CT\ =\ \frac{\hb}{2\pi}\int_{-\infty}^{+\infty} dx\ \d_x\Phi\ =\
\frac{\hb}{2\pi}\ \Bigl( \Phi(x=\infty)-\Phi(x=-\infty) \Bigr) .  }
The topological solitons that correspond to single particles in the
quantum theory are described classically by field configurations with
$\CT=\pm1$. These solitons are kinks that connect two neighboring
vacua in the $\cos(\hb\Phi)$ potential.

The sine-Gordon model possesses infinitely many local integrals of motion
with odd Lorentz spins, we denote them by $\CI_n$. Besides those,
the sine-Gordon model also admits four non-local conserved
currents \ref\rBL{D. Bernard and A. Leclair, {\it "Quantum Group Symmetries
and Non-Local Currents in 2D QFT"}, to appear in Comm. Math. Phys.}:
\eqn\IIIxi{\d_{\mu}\ J_{\mu}^{\pm}(x,t)\ =\
\d_{\mu}{\bar J}_{\mu}^{\pm}(x,t)\ =\ 0.}
The Lorentz spin $s$ of the currents $J_\mu^\pm$ $\({\bar J}_\mu^\pm\)$
are $s=\frac{2}{\hb^2}$ $\(-\frac{2}{\hb^2}\)$.
{}From these conserved currents we define four conserved charges,
$Q_{\pm}$ and $\bQ_{\pm}$, respectively associated to the
currents $J^{\pm}_{\mu}(x,t)$ and ${\bar J}_{\mu}^{\pm}(x,t)$.
The Lorentz spins of the conserved charges are :
\eqn\IIIxiv{ {\rm spin}\(Q_\pm\)\ =\
-{\rm spin}\({\bar Q}_\pm\)\ =\
\frac{2-\hb^2}{\hb^2}= \frac{8\pi -\beta^2}{\beta^2} .}
The conserved currents whose exact expressions are given in ref. \rBL\
are Mandelstam like  vertex operators \ref\RMan{S. Mandelstam,
Phys. Rev. D11 (1976) 3026}
and are thus non-local.

The algebra of the non-local charges is :
\eqna\IIIxxiii  $$\eqalignno{
Q_{\pm}\ \bQ_{\pm} - q^{2}\ \bQ_{\pm}\ Q_{\pm} \ &=\ 0 &\IIIxxiii a\cr
Q_{\pm}\ \bQ_{\mp} - q^{-2}\ \bQ_{\mp}\ Q_{\pm} \ &=\
a \BL 1- q^{\pm 2\CT}\BR
&\IIIxxiii b\cr
\BBL\ \CT\ ,\ Q_{\pm}\ \BBR\ &=\ \pm2\ Q_{\pm} &\IIIxxiii c\cr
\BBL\ \CT\ ,\ \bQ_{\pm}\ \BBR\ &=\ \pm2\ \bQ_{\pm} , &\IIIxxiii d\cr}$$
where $q=\exp(-2\pi i /\hb^2)$.
and $a$ some constant.
The algebra \IIIxxiii{}\ is a known infinite
dimensional algebra, namely the q-deformation of the $sl(2)$
affine Kac-Moody algebra, denoted $\slq$, with zero center
\ref\rDrin{V. Drinfel'd, Sov. Math. Dokl. 32 (1985) 254; Sov. Math. Dokl.
36 (1988) 212}
\ref\rJim{M. Jimbo. Lett. Math. Phys. 10 (1985) 63; Lett. Math. Phys.
11 (1986) 247}.
Only the Serre relations for
$\slq$ are missing in \IIIxxiii{}.

The non-local charges \IIIxi{}\ provide relevent information;
for example, the S-matrix of the Sine-Gordon solitons
\ref\rZZ{A. Zamolodchikov and Al. Zamolodchikov, Annals Phys.
120 (1979) 253} can be deduced
from this $\slq$ symmetry plus its unitary and crossing symmetry property.
However, they probably do not form a complete symmetry algebra
because the local integrals of motion do not seem to be generated
by them. To prove the conjecture that the local conserved charges
$\CI_n$ and these non-local charges generate a complete symmetry
algebra for the sine-Gordon models will be very illuminating.

\newsec{ QUANTUM SYMMETRIES IN 2D LATTICE FIELD THEORY.}

We consider vertex models, i.e. models of two-dimensional statistical
mechanics in which the discrete spin variables live on the midpoints of the
links of a square lattice, and the Boltzmann weights are
associated to the vertices of the lattice \ref\Vertex{Cf e.g. R.J. Baxter,
{\it Exactly solved models in statistical mechanics}, Academic Press 1982.}.
The Boltzmann weight
of a given vertex depends on the spin variables $\sigma_1,
\dots,\sigma_4$ at the four
sites surrounding the vertex, and is denoted by
$R_{\sigma_1\sigma_2}^{\sigma_3\sigma_4}$.
It is useful to view $R$ as an operator
$V\otimes V\to V\otimes V$, where $V$ is the vector space spanned
by a set of basis vectors $e_\sigma$ labeled by
the possible values of the spin variable:
\eqn\eai{
R e_{\sigma_1}\otimes e_{\sigma_2}=R_{\sigma_1\sigma_2}^{\sigma_3\sigma_4}
\  e_{\sigma_3}\otimes e_{\sigma_4}.}

Partition function and correlation functions are defined as follows.
Consider the system in a finite square box of size $N\times N$. Then the
square lattice $\La = \Z^2 \big/ N^2\Z^2$ contains points with integer
coordinates $(x,t)$ called space and time. The spin variables live on the
lattice $\La'$ of points of the form $(x+\half,t)$ and $(x,t+\half)$
with $x,\ t$ integers modulo $N$. For each $i\in\La'$ introduce a copy
$V_i$ of the space $V$. For $(x,t)\in \La$ define $R(x,t)$ be the matrix
$R$ mapping $V_{(x-\half,t)}\oti V_{(x,t-\half)}$ to
$V_{(x+\half,t)}\oti V_{(x,t+\half)}$. The matrix of Boltzmann weights,
\eqn\eaii{
B\ =\ \bigotimes_{(x,t)\in\La}\ R(x,t),}
is an operator from $\bigotimes_{i\in\La'}V_i$ to $\bigotimes_{j\in\La'} V_j$.

The partition function is $Z_N\ =\ {\rm tr}\ B$.

For any operator $\CO\in{\rm End}V$ define the insertion $\CO(j)$
of $\CO$ at the point $j\in\La'$ to be the operator
$1\oti\cdots\oti1\oti\CO\oti 1\oti\cdots\oti1$ acting on $V_j$
in the tensor product $\bigotimes_{i\in\La'}V_i$. The correlation
functions of operator insertions are defined as
\eqn\eaiii{
\vev{\CO_1(j_1)\ \cdots\ \CO_n(j_n)}_N\ =\
\inv{Z_N}\ {\rm tr}\BL\ \prod_{k=1}^n\CO_k(j_k)\ B\ \BR.}
Classical examples are $\CO e_{\sig}=\sig e_{\sig}$ and
$\CO e_{\sig}=\de_{\sig {\bar \sig}}e_{\sig}$ for some ${\bar \sig}$.
Their correlation functions are the usual spin correlation functions
and the joint probabilities that the spin $\sig_{j_k}$ assume given values.

An alternative formalism is the transfer matrix formulation. In this
formalism, one assigns to each $x=1,\cdots,N$ a copy $V_x$ of $V$.
The transfer matrix $T$ is:
\eqn\eaiv{
T\ =\ {\rm tr}_0\BL\ R_{0N}\cdots R_{02}\ R_{01}\ \BR}
with $R_{nm}$ the matrix $R$ acting on $V_n$ and $V_m$ in
$V_0\oti V_1\oti\cdots\oti V_N$, and the trace is over $V_0$.

For $\CO\in{\rm End}V$ and $x\in\{1,\cdots,N\}$ define
$\CO(x)$ as $\CO$ acting on $V_x$ in $V_1\oti\cdots\oti V_N$
and $\CO(x-\half)$ as
$\CO(x-\half)=T^{-1}{\rm tr}_0\BL R_{0N}\cdots R_{0x}\
\CO_0\ R_{0,x-1}\cdots R_{01}\BR$. Heisenberg fields $\CO(j)$,
$j\in\La'$ are defined as:
\eqn\eav{\eqalign{
\CO(x-\half,t)\ &=\ T^{-t}\ \CO(x-\half)\ T^t \cr
\CO(x,t-\half)\ &=\ T^{-t}\ \CO(x)\ T^t \cr}}
The partition function in the transfer matrix formalism is
$Z={\rm tr} T^N$, and if the time coordinates of $j_1,\cdots,j_n$
are ordered (with smaller times on the right of larger times)
the correlation function $\vev{\CO_1(j_1)\cdots\CO_n(j_n)}_N$
defined above coincides with
\eqn\eavi{
\vev{\CO_1(j_1)\cdots\CO_n(j_n)}_N\ =\
\inv{Z_N}\ {\rm tr}\BL\
\CO_1(j_1)\cdots\CO_n(j_n)\ T^N\BR}

We have defined point-like operator insertions in two different
formalisms. It is sometimes useful to define operator insertions
associated to some finite set of neighboring points in $\La'$
as linear combination of products of point-like insertions
at the points of the set. This is the lattice analogue of the
operator product expansion of field theory.

\bigskip
\noindent{\bf 1a) Quantum symmetries and conserved currents.}

\noindent{\bf (i) Lie algebra symmetry. }
Suppose that the Boltzmann weights are invariant under some
Lie algebra $\CG$.  This means that $V$ carries a representation
of $\CG$ and for each generator $T_a$ of $\CG$ in that representation
we have
\eqn\eaai{
R(T_a\otimes 1 +1\otimes T_a)=(T_a\otimes 1+ 1 \otimes T_a)R}
i.e. $R$ is an intertwiner. It is useful to represent this
equation graphically. If $T_a$ is represented by a little
cross, we have
\eqn\eaaii{
\matrix{~&\Big\vert& ~\cr -& -- & - \cr
		~&\Big\vert& ~\cr}\ +\
\matrix{~&\Big\vert& ~\cr -& -- & - \cr
		~&\Big\vert& ~\cr}\ =\
\matrix{~&\Big\vert& ~\cr -& -- & - \cr
		~&\Big\vert& ~\cr}\ +\
\matrix{~&\Big\vert& ~\cr -& -- & - \cr
		~&\Big\vert& ~\cr} }
Introduce now a local current $J^\mu(x,t;X)$, linear in $X\in \g$,
for each vertex $(x,t)$ of the lattice. The components $J^t(x,t;X)$,
$J^x(x,t;X)$ are defined by the insertion of the matrix $X=\sum_a X_aT_a$
at the site $(x,t-{1\over2})$ or the site $(x-{1\over2},t)$,
respectively. Graphically,
\eqna\eaaiii
$$\eqalignno{
 J^t(x,t;X)\ &=\
\matrix{~&\Big\vert& ~\cr -& -- & - \cr
		~&\Big\vert& ~\cr} {}^{(x,t)}
&\eaaiii a\cr
 ~ &~ \cr
 J^x(x,t;X)\ &=\
\matrix{~&\Big\vert& ~\cr -& -- & - \cr
		~&\Big\vert& ~\cr} {}^{(x,t)}
&\eaaiii b\cr }$$
Then, in any correlation function (with no insertion of other fields
at the sites surrounding $(x,t)$), \eaai\ reads
\eqn\eaaiv{
J^t(x,t+1;X)-J^t(x,t;X)+J^x(x+1,t;X)-J^x(x,t;X)=0}
which is the lattice version of the continuity equation
$\partial_\mu J^\mu=0$. As in the continuum, this equation implies
the conservation of the charge $Q(X)=\sum_x J^t(x,t;X)$.

As it is obvious from the pictures \eaaiii{},
the operators $J^\mu(x;X)$ are local operators: they
satisfy equal-time commutation relations:
\eqn\eaaivb{
J^\mu(x;X)\ J^\nu(y;Y) \ =\
J^\nu(y;Y)\ J^\mu(x;X) \quad ;\quad \forall\ x\not= y }
for all $X,Y\in\CG$.

\noindent {\bf (ii) Quantum invariance.}
We now generalize \ref\rBgio{D. Bernard and G. Felder, {\it Quantum
group symmetries in 2D lattice quantum field theory}, to appear
in Nucl. Phys. B}
the preceding construction to an invariance
under a Hopf algebra. Recall that a Hopf algebra $A$ is an algebra with
unit $1$ and associative product $m:A\x A\to A$, equipped with a coproduct
$\De:A\to A\x A$, a counit $\ep:A\to \C$, and an antipode
$S: A\to A$ so that: (i) $\De$, $\ep$ are algebra homomorphisms,
$S$ is an algebra antihomomorphism; (ii) $(1\x\De)\De(X)=(\De\x1)\De(X)$;
(iii) $(1\x\ep)\De(X)=(\ep\x1)\De(X)=X$; (iv) $m(1\x S)\De(X)=
m(S\x 1)\De(X)=\ep(X)1$, for all $X\in A$.
The Hopf algebra $A$ we will consider are those
generated by elements $T_a$, $\Theta_a^b$,
${\hat \Th}^b_a$ with, among the relations, $\Th^c_a{\hat \Th}^b_c
={\hat \Th}^c_a\Th^b_c=\de^b_a$.
We also assume that the comultiplication in $A$ is defined by:
\eqn\eaav{\eqalign{
\De(T_a)\ &=\ T_a\x 1 + \Th_a^b\x T_b \cr
\De(\Th^b_a)\ &=\ \Th_a^c\x\Th^b_c \cr
\De(\hat \Th^a_b)\ &=\ \hat \Th_b^c\x\hat \Th^a_c \cr}}
The definition of the counit and the antipode  in $A$
are found from the Hopf algebra axioms.
The motivation for introducing these algebras will be given latter.
Lie superalgebras provide an example of such algebras.

The correct generalization of the invariance eq. \eaai\ is
$R\ \Delta(X)=\sigma\circ\Delta(X)\ R,$
with $\sigma X\otimes Y=Y\otimes X$. Explicitly,
\eqna\eaavii
$$\eqalignno{
R\ (T_a\otimes 1+\Theta_a^b\otimes T_b) &=
(1\otimes T_a+T_b\otimes \Theta_a^b)\ R &\eaavii a \cr
R\ \Theta_a^c\otimes\Theta_c^b &=\Theta_c^b\otimes\Theta_a^c\ R
&\eaavii b\cr}$$
These equations have a graphical interpretation. The generators
$T_a$, $\Theta_a^b$, $\hat\Theta_a^b$ are conveniently represented
in terms of crosses and oriented wavy lines:
$$
           T_a={}^a\ \ \times \qquad;\qquad
    \Theta_a^b={}^a\ \ \ \ \ \ \ \ \ \ {}^b\qquad;\qquad
\hat\Theta_a^b={}^b\ \ \ \ \ \ \ \ \ \ {}^a
$$
The graphical representation of \eaavii{a}\ is then
$$
\matrix{~&\Big\vert& ~\cr -& -- & - \cr
		~&\Big\vert& ~\cr}\ +\
\matrix{~&\Big\vert& ~\cr -& -- & - \cr
		~&\Big\vert& ~\cr}\ =\
\matrix{~&\Big\vert& ~\cr -& -- & - \cr
		~&\Big\vert& ~\cr}\ +\
\matrix{~&\Big\vert& ~\cr -& -- & - \cr
		~&\Big\vert& ~\cr}
$$
with the convention that where pieces of wavy lines join
an implicit contraction of indices is understood.
The currents $J^\mu(x,t;X)$, $X=\sum_a X_aT_a$,
are then constructed as for parafermionic currents,
namely with a disorder line (the wavy line) attached:
\eqna\eaaviii
$$\eqalignno{
J_a^t(x,t)=J^t(x,t;T_a)&= \matrix{~\cr ~\cr {}_a\cr}
\matrix{~&\Big\vert& ~\cr -& - & - \cr
		~&\Big\vert& ~\cr}
\matrix{~&\Big\vert& ~\cr -& - & - \cr
		~&\Big\vert& ~\cr}\
\matrix{\cdots \cr \cdots \cr \cdots\cr}\
\matrix{~&\Big\vert& ~\cr -& - & - \cr
		~&\Big\vert& ~\cr}
\matrix{~&\Big\vert&{}_{(x,t)}\cr -& - & - \cr
		~&\Big\vert& ~\cr}\cr
     ~&~ &\eaaviii a\cr
J_a^x(x,t)=J^x(x,t;T_a)&= \matrix{{}^a\cr ~\cr ~\cr}
\matrix{~&\Big\vert& ~\cr -& - & - \cr
		~&\Big\vert& ~\cr}
\matrix{~&\Big\vert& ~\cr -& - & - \cr
		~&\Big\vert& ~\cr}\
\matrix{\cdots \cr \cdots \cr \cdots\cr}\
\matrix{~&\Big\vert& ~\cr -& - & - \cr
		~&\Big\vert& ~\cr}
\matrix{~&\Big\vert& {}_{(x,t)}\cr -& - & - \cr
		~&\Big\vert& ~\cr}\cr
 ~&~ &\eaaviii b\cr}$$
The disorder line ends at some specified
point on the boundary of the lattice.
The identity \eaavii{b}\ implies that the disorder line may be deformed
(away from insertions of observables)
without changing the value of correlation functions. It
behaves as the holonomy of a flat connection, just as
for ordinary disorder fields \ref\Disorder{L.P. Kadanoff
and H. Ceva, Phys. Rev. B3 (1971) 3918\semi
E. Fradkin and L.P. Kadanoff, Nucl. Phys. B170 (1980) 1\semi
J. Fr\"ohlich and P. A. Marchetti,
Comm. Math. Phys. 112 (1987) 343.}.
Equation \eaavii{a}\ implies the
continuity equation \eaaiv\ for non-local currents.

In the operator formalism, the time component of the current is
an operator (in the Schr\"odinger picture) acting on the finite volume
Hilbert space $V\otimes\cdots\otimes V$ ($N$ factors) as
\eqn\eaaix{
J^t_a(x)=J^t(x;T_a)=
\Theta_a^{a_1}\otimes\Theta_{a_1}^{a_2}\otimes\cdots\otimes
\Theta_{a_{x-2}}^{a_{x-1}}\otimes T_{a_{x-1}}\otimes
1\otimes\cdots\otimes 1.}
The space component has a more cumbersome operator representation
which we omit.

\bigskip
\noindent {\bf (iii) The braiding relations.} By construction
the currents \eaaviii{}\ are non-local. They satisfy
braided equal-time commutation relations. These braiding relations
arise due to the topological obstructions that one encounters when trying
to move the wavy string attached to the currents through a point
on which a field is located.
In order to write simple closed formula for the braiding
relations we now assume that we have completed the set of generators,
$T_a$, $\Th^b_a$, $\hat \Th^b_a$ such that they close under the
adjoint action. This implies that there exists a c-number matrix
$R^{bd}_{ac}$ such that:
\eqn\eaax{\eqalign{
\Th_a^n\ T_c\ \hat \Th^b_n\
&=\ R^{bd}_{ac}\ T_d \cr
R^{ab}_{nm}\ \Th^n_c\ \Th^m_d\ &=\ \Th^b_m\ \Th^a_n\ R^{nm}_{cd}\cr}}
Then, a simple computation shows that:
\eqn\eaaxi{J^\mu_a(x)\ J^\nu_b(y)\ = R^{cd}_{ab}\ J^\nu_d(y)
J_c^\mu(x),\qquad\qquad {\rm for}\ x>y }

\bigskip
\noindent {\bf (iii) Global symmetry algebra.} The algebra $A$ acts
on the Hilbert space $V^{\otimes N}$ by the coproduct $\Delta_N$,
defined recursively by $\Delta_2=\Delta$, $\Delta_{n+1}=
\Delta_n(1\otimes\Delta)$. For generators, we have the formulae
\eqn\eaax{
\eqalign{
\Delta_N(\Theta_a^b)&=\Theta_a^{a_1}\otimes\Theta_{a_1}^{a_2}\otimes
\cdots\otimes\Theta_{a_{N-1}}^b\cr
\Delta_N(\hat\Theta_a^b)
&=\hat\Theta^b_{a_1}\otimes\hat\Theta^{a_1}_{a_2}\otimes
\cdots\otimes\hat\Theta^{a_{N-1}}_a\cr
\Delta_N(T_a)&=\sum_{x=1}^N
\Theta_a^{a_1}\otimes\Theta_{a_1}^{a_2}\otimes\cdots\otimes
\Theta_{a_{x-2}}^{a_{x-1}}\otimes T_{a_{x-1}}\otimes
1\otimes\cdots\otimes 1.\cr}}
Comparing with \eaaix, we see that $\Delta_N(T_a)$ is the charge
corresponding to the current $J^\mu_a(x)$:
\eqn\eaaxi{
\Delta_N(T_a)=\sum_{x=1}^NJ^t_a(x).}
The global charges $\De_N(\Th^b_a)$ can be interpreted as topological charges.

The global charges $\De_N(T_a)$ and $\De_N(\Th^b_a)$,
satisfy the same algebra as
the original generators $T_a$, $\Th^b_a$ (because the comultiplication
$\De$ is an homorphism from $A$ to $A\x A$). If we
assume, as we did in the previous section, that the generators
$T_a$, $\Th^b_a$ are closed under the adjoint action, then there
also exist c-numbers $f^a_{bc}$ such that :
\eqn\eaaxii{ T_a T_b - R^{cd}_{ab}\ T_dT_c\ =\ f^c_{ab}\ T_c.}
These are generalized braided Lie commutation relations.

\noindent {\bf Remark:} Because the currents are non-local, the local
conservation laws for the currents do not systematically imply
those of the global charges. The conservation laws for the charges
can be broken by boundary terms which depends on the sector on which
the charges are acting.

\bigskip
\noindent{\bf 1b)  Fields multiplets.}

The symmetry algebra $A$ acts also on the field operators.
In the operator formalism, the fields are operators $\CO\in
{\rm End}(V^{\x^N})$.
Any element $X\in A$ of an Hopf algebra $A$ acts an operator $\CO$
by
\eqn\eadj{ Q_X\ \CO= \sum_i\ X_i\ \CO\ S(X^i) }
with $\De(X)=\sum_i X_i\x X^i$
and $S$ the antipode in $A$. In our case, for the generators
$T_a$ and $\Th_a^b$, this becomes:
\eqn\eabi{\eqalign{
Q_a\CO\ &=\ \De_N(T_a)\ \CO\ - (Q_a^b\CO)\ \De_N(T_b) \cr
Q_a^b\CO\ &=\ \De_N(\Th^c_a)\ \CO\ \De_N({\hat \Th^b_c}) \cr}}

The multiplets of fields are collections of fields transforming
in a representation of the symmetry algebra $A$.
More precisely, let $V_\Lambda$ be
a representation space for $A$. A field multiplet at $x$ is an operator
$\Phi^\La(x;v)$ acting on  $V^{\otimes N}$ depending linearly on
a vector $v$ in $V_\Lambda$, with transformation property
\eqn\eabii{
 Q_X\Phi^\Lambda(x;v)=\Phi^\Lambda(x;Xv).}
It is clear that, in general, the fields $\Phi^\Lambda(x;v)$ are
necessarily non-local. However, in a given multiplet there
can be local fields.

Field multiplets can be constructed from the following data:
Representation spaces $W$, $W'$ and operators $\phi$ and $\Om$,
$\phi :V_\Lambda\otimes W \to W'$ and
$\Omega : V_\Lambda\otimes V\to V_\Lambda\otimes V$,
with (twisted) intertwining properties
\eqn\eabiii{\eqalign{
\phi\Delta(X)&=X\phi\cr
\Omega\Delta(X)&=\sig\circ\Delta(X)\Omega,\cr}}
for all $X\in A$.
The spaces $W$, $W'$ are in the simplest case equal to
$V$, or may be tensor products $V^{\otimes n}$, $V^{\otimes n'}$. In terms of a
basis $\{e_i\}$ of $V_\Lambda$, with $\phi(e_i\otimes w)=\phi_iw$,
$\Omega(e_i\otimes v)=e_j\otimes \Omega_i^jv$, $Xe_i=e_jX_i^j$,
the field multiplets are operators acting on
$V^{\otimes N}$ defined by,
\eqn\eabiv{
\Phi^\La(x;e_i)\equiv \Phi_i^\La(x)=
\Omega_i^{i_1}\otimes\Omega_{i_1}^{i_2}\otimes
\cdots\otimes\Omega_{i_{x-2}}^{i_{x-1}}
\otimes\phi_{i_{x-1}}\otimes1\otimes\cdots\otimes1,}
with $\phi_{i_x}$ acting on the tensor product of the $x$th to
the $(x-1+n)$th factor \footnote*{If $n\not=n'$ the insertion
of a field produces a deformation of the lattice. It is understood
that we consider correlation functions where the total $n$ is equal
to the total $n'$ so that at infinity the lattice is the regular
square lattice.}.
Graphically this is represented as
\def\grapphi{
\matrix{~&\Big\vert& ~\cr =& == & = \cr
		~&\Big\vert& ~\cr}
\matrix{\Big\vert\cr \cr \Big\vert\cr}
\matrix{~&\Big\vert& ~\cr =& == & = \cr
		~&\Big\vert& ~\cr}
\matrix{\cdots \cr \cdots \cr \cdots\cr}
\matrix{~&\Big\vert& ~\cr =& == & = \cr
		~&\Big\vert& ~\cr}
\matrix{\Big\vert\cr \bigcirc  \cr
	\Big\vert \cr} \ \
\matrix{ ~&\Big\vert \cr ~&\cdot \cr ~&\Big\vert \cr}\ \
\matrix{ ~&\Big\vert \cr ~&\cdot \cr ~&\Big\vert \cr} }
$$
\Phi^\La_i(x)\ \equiv\ {}_i\ \grapphi
$$
\noindent where the circle represents an insertion of $\phi$
and each crossing of the double line with a single line
represents an $\Omega$.

The transformation property \eabii\ follows from \eabiii\ and Hopf
algebra properties.
The action of the algebra of fields is by definition \eabi :
\eqn\eabvi{ \eqalign{
 Q^b_a\Phi^\La(x;v)&=
\Theta_a^{a_1}\otimes\Theta_{a_1}^{a_2}\otimes\cdots\otimes
\Theta_{a_{N-1}}^c\
\Phi^\La(x;v)\
\hat\Theta_{b_1}^b\otimes\hat\Theta_{b_2}^{b_1}\otimes\cdots
\otimes\hat\Theta_c^{b_{N-1}},\cr
 Q_a\Phi^\La(x;v)&=\sum_{y=1}^N
\BL J_a^t(y)\ \Phi^\La(x;v)-Q_a^b\Phi^\La(x;v)\ J_b^t(y)\BR.\cr}}
Because the terms with $y>x+n-1$ cancel in the sum,
the action of the charges $Q_a$ can written as a contour integral on
the lattice.
Indicating graphically the summation by an integration contour on
the dual lattice, we have:
\eqn\eabvi{
 Q_a\Phi_i^\La(x)\ \equiv\ \matrix{ {}^a\cr {}_i\cr ~\cr}\ \grapphi}
The integration contour is surrounding the fields.
The intertwining properties \eabiii\ of the microscopic data $\phi$ and
$\Om$ imply that the field multiplets defined in \eabiv\
transform covariantly:
\eqn\eabbis{\eqalign{
Q^b_a\ \Phi^\La_i(x)\ =\ \Th^{bj}_{ai} \Phi^\La_j(x)\cr
Q_a\ \Phi^\La_i(x)\ =\ T^j_{ai} \Phi^\La_j(x)\cr}}

\noindent {\bf Remark:} Because the field multiplets are non-local
they satisfy equal-time braiding relations. But
the braiding relations between the field multiplets are constrained
by the quantum invariance. Let $V_{\La}$ and $V_{\La'}$ be two
representation spaces of $A$ with basis $e_i\in V_{\La}$ and
$e_{\al}'\in V_{\La'}$. Let $\Phi_i(x)\equiv\Phi(x;e_i)$
and $\Phi_{\al}'\equiv\Phi'(y;e_{\al}')$ be two field multiplets.
Denote by $\CR:V_{\La'}\oti V_{\La}\to V_{\La'}\oti V_{\La}$, with
$\CR(e_{\al}'\oti e_i)=\CR_{\al i}^{\beta j}\ e_{\beta}'\oti e_j$,
the braiding matrix of these field multiplets:
\eqn\eabvii{\Phi_i(x)\Phi_{\al}'(y)\ =\ \CR^{\beta j}_{\al i}\
\Phi_{\beta}'(y)\Phi_j(x)
\qquad ~~~~~~~ {\rm for}\ x>y}
By quantum invariance,
\eqn\eabviii{ \CR\ \De(X)\ =\ \sig\circ\De(X)\ \CR
\qquad ,\qquad \forall X\in A}
Thus, the braiding matrices intertwines the quantum algebra.
Examples will be given in the following sections.

\bigskip
\noindent {\bf 1c) Examples.}

\noindent {\bf (i) Yangian invariance.}
The Yangians are deformations of loop algebras which have
been introduced by Drinfel'd \rDrin. They are related to rational
solutions of the quantum Yang-Baxter equation.

Let us first recall what are the Yangians. Let $\CG$ be a simple Lie
algebra with structure constants $f_{abc}$ in an orthonormalized
basis. The Yangian, denoted $Y(\CG)$, is the associative algebra with
unity generated by the elements $t_a$ and $\CT_a$, $a=1,\cdots,
{\rm dim}\CG$, satisfying the relations:
\eqn\eaci{\eqalign{
\BBL\ t_a\ ,\ t_b\ \BBR\ &=\ f_{abc}\ t_c \cr
\BBL\ t_a\ ,\ \CT_b\ \BBR\ &=\ f_{abc}\ \CT_c \cr
\BBL \CT_a,\BBL\CT_b, t_c \BBR\BBR \ &-\
\BBL t_a,\BBL\CT_b, \CT_c \BBR\BBR \
=\  A_{abc}^{lmn}\ \Big\{ t_l, t_m, t_n \Big\} \cr} }
with $A_{abc}^{def}=\frac{1}{24} f_{adk}f_{bel}f_{cfm}f^{klm}$
and $\{x_1,x_2,x_3\}=\sum_{i\not= j\not= k}x_ix_jx_k$.
In particular, the elements $t_a$ generate the Lie
algebra $\CG$ and the elements $\CT_a$ are $\CG$ - intertwiners taking
values in the adjoint representation of $\CG$.
(for $\CG=SU(2)$ one must add another Serre-like relation.)
The Yangians $Y(\CG)$ are
Hopf algebras with comultiplication $\De$, counit $\ep$ and
antipode $S$ defined by:

\eqna\eacii
$$\eqalignno{
\De\ t_a\ &=\ t_a\oti 1 + 1\oti t_a\ ; &\eacii c\cr
\ep(t_a)\ &=\ 0 \quad ;\quad S(t_a)\ =\ - t_a \cr
\De\ \CT_a\ &=\ \CT_a\oti 1 + 1\oti \CT_a
- \half f_{abc}\ t_b\oti t_c\ ;&\eacii b \cr
\ep(\CT_a)\ &=\ 0 \quad ;\quad
S(\CT_a)\ =\ - \CT_a - \frac{C_{\rm Ad}}{4} t_a \cr}$$
with $C_{{\rm Ad}}$ the Casimir in the adjoint representation of $\CG$:
$f_{abc}f_{bcd}=C_{\rm Ad} \ \de_{ad}$.

The Yangian invariant $R$ - matrices are those which satisfy the
intertwining relation
$R\ \De(Y)\ =\ \sig\circ\De(Y)\ R$
for all $Y\in Y(\CG)$.
We suppose that the vertex models we are considering
in this section are defined
from $Y(\CG)$ - invariant Boltzmann weights.
The non-local conserved currents we will describe in this section
are the lattice analogues of those hidden in 2D massive current
algebras \ref\Lu {M. L\"uscher, Nucl. Phys. B135 (1978) 1.}
\ref\denis {D. Bernard,
Comm. Math. Phys. 137 (1991) 191}, see section 4.

{}From the defining relations of $Y(\CG)$, it is obvious
that the Yangians $Y(\CG)$ possess the properties we need in order
to apply our formalism.
We therefore can define Yangian currents. For simplicity, we just define the
currents associated to the generators $t_a$ and $\CT_a$; we denote
them $J_a^{\mu}(x,t)$ and $\CJ_a^{\mu}(x,t)$, respectively.
By applying the general construction,
we deduce that the conserved currents $J_a^{\mu}(x,t)$
are local (as it should be);
they are defined by local insertions of the matrices $t_a$ representing
the Lie algebra $\CG$. The conserved currents $\CJ_a^{\mu}(x,t)$ are non-local;
in the operator formalism we have:
\eqn\eaciii{
\CJ_a^{\mu}(x,t)\ =\ \CT_a^{\mu}(x,t) + \half f_{abc}\
J_b^{\mu}(x,t)\ \phi_c(x,t)}
with
\eqn\eaciv{
\phi_c(x,t)\ =\ \sum_{y<x}\ J_c^t(y,t)}
In eq. \eaciii\ and \eaciv, the notations $J_a^{\mu}(x,t)$ and
$\CT_a^{\mu}(x,t)$ refer to insertions of the matrices
$t_a$ or $\CT_a$ on the link oriented in the direction $\mu$
and ending at the point $(x,t)$. Notice the similarity
between the expression of these lattice non-local currents
and of their continuous partners \Lu\ \denis\ and section 4b.

The braiding relations for the currents $J_a^{\mu}(x,t)$ and
$\CT_a^{\mu}(y,t)$ follows from our previous discussion:
\eqn\eacv{\eqalign{
J_a^{\mu}(x,t)\ J_b^{\nu}(y,t)\ &=\
J_b^{\nu}(y,t)\ J_a^{\mu}(x,t)\
 ~~~~~~~~~~~~~~~~~~~~~~~~~~~~~~;\ \forall\ x \not= y \cr
\CJ_a^{\mu}(x,t)\ J_b^{\nu}(y,t)\ &=\
J_b^{\nu}(y,t)\ \CJ_a^{\mu}(x,t)\
 ~~~~~~~~~~~~~~~~~~~~~~~~~~~~~~;\ {\rm for}\ x < y \cr
\CJ_a^{\mu}(x,t)\ J_b^{\nu}(y,t)\ &=\
J_b^{\nu}(y,t)\ \CJ_a^{\mu}(x,t)\ \cr
&~\quad -\half f_{anm} \BL f_{nbk} J_k^{\nu}(y,t)\BR\ J_m^{\mu}(x,t)
 ~~~~~~~~~;\ {\rm for}\ x > y \cr}}

\bigskip
\noindent{\bf (ii)  Quantum universal enveloping algebras.}
For any (affine) Kac-Moody algebra $\g$ with Cartan matrix
$a_{ij}$, $0\leq i,j\leq r$ and any complex number $q\not=0$,
Drinfel'd \rDrin\
and Jimbo \rJim
define an universal
quantum enveloping algebra (QUEA) $U_q(\g)$. Let $d_i$ be positive
integers such that the matrix $d_ia_{ij}$ is symmetric, and let
$q_i=q^{d_i}$. The algebra $A=U_q(\g)$ has generators
$E^+_i$, $E^-_i$, $K_i^2$, $K_i^{-2}$, $0\leq i\leq r$, and relations
\eqn\eacvi{\eqalign{
K_i^2K_j^{\pm 2}&=K_j^{\pm 2}K_i^2,\cr
K_i^2K_i^{-2}&=K_i^{-2}K_i^2=1,\cr
K_i^2E_j^\pm&=q_i^{\pm a_{ij}}E^\pm_jK_i^2,\cr
E_i^+E_j^--q_i^{-a_{ij}}E^-_jE^+_i&=\delta_{ij}(K_i^4-1),\cr}}
plus Chevalley-Serre relations to be written below. The Hopf
algebra structure is defined by the coproduct
\eqn\eacvi{\eqalign{
\Delta(K_i^{\pm 2})&=K_i^{\pm 2}\otimes K_i^{\pm 2}, \cr
\Delta(E^\pm_i)&=E^\pm_i\otimes 1+ K_i^2\otimes E_i^\pm, \cr}}
counit $\epsilon(E^\pm_i)=0$, $\epsilon(K_i^{\pm 2})=1$, and
antipode $S(E^\pm_i)=-K_i^{-2}E_i^\pm$, $S(K_i^{\pm 2})=K^{\mp2}_i$.
The adjoint representation is then defined as usual, eq. \eadj, and
the Chevalley-Serre relations are
\eqn\eacvii{
\Ad_{E_i^\pm}^{1-a_{ij}}E_j^\pm=0.}
We see that this is a very simple example of the Hopf
algebras described in the introduction: the
generators $T_a$ are $E_i^\pm$ and $\Theta_a^b$ is
diagonal with entries $K_i^2$.

Statistical models with QUEA symmetry are defined by
trigonometric solutions of the Yang-Baxter equation, the
simplest case being the six-vertex model \Vertex.

The simple currents are defined by insertions of $E^\pm_i$
with disorder lines given by insertions of $K_i^2$.
In the operator formalism, the time components of the currents
are
\eqn\eacviii{
J_i^{t\pm}(x)=K_i^2\otimes\cdots\otimes K_i^2\otimes E_i^\pm
\otimes 1\otimes\cdots\otimes 1.}
The corresponding charges are the generators $\Delta_N(E_i^\pm)$
acting on the whole space $V^{\otimes N}$. The braiding
relations are, for $x>y$:
\eqn\eacix{\eqalign{
J_i^{\mu\pm}(x)
J_j^{\nu\pm}(y)&=q_i^{\pm a_{ij}}
J_j^{\nu\pm}(y)
J_i^{\mu\pm}(x),\cr
J_i^{\mu\pm}(x)
J_j^{\nu\mp}(y)&=q_i^{\mp a_{ij}}
J_j^{\nu\mp}(y)
J_i^{\mu\pm}(x).\cr}}
These relations are the same as the braiding relations of
chiral vertex operators of a free massless field $\phi$
taking value in the Cartan subalgebra of $\g$ with
canonical momentum $\pi$. This suggests
the continuum limit identification
\eqn\eacx{
J_j^{t\pm}\sim\exp\( i\beta\alpha_j\(\pm\phi(x)+
\int_{-\infty}^x\pi(y)dy\)\) ,}
with $q=e^{i\beta^2}$, $\alpha_j$ the simple roots, with inner
product $\alpha_i\alpha_j=d_ia_{ij}$.  The space component
in the continuum limit is $J^{x\pm}_j=\mp i J^{t\pm}_j$.

\newsec{CLASSICAL ORIGIN OF QUANTUM SYMMETRIES: DRESSING TRANSFORMATIONS.}

The dressing transformations form the (hidden) symmetry
groups of solitons equations. Dressing transformations were first
introduced by V. Zakharav and A. Shabat \ref\rZS{V. Zakharov and A. Shabat,
Funct. Anal. 13 (1979) 166}
and futher developped by the Kyoto group in their Tau-function appraoch
to soliton equations \ref\rKyoto{E. Date, M. Jimbo, M. Kashiwara and
T. Miwa, Proc. Japan Acad. 57A (1981) 342;
ibid. 57A (1981) 387; J. Phys. Soc. Japan 50 (1981) 3806;
Physica 4D (1982) 343; Publ. RIMS 18 (1982) 1111; J. Phys. Soc. Japan
50 (1981) 3813}.
Their Poisson structure was disantangled by
M. Semenov-Tian-Shansky \ref\rSem{M. Semenov-Tian-Shansky, Funct. Anal.
Appl. 17 (1983) 259; Publ. RIMS 21 (1983) 1237}.
The author's understanding of these transformations
emerged from a joint work with O. Babelon \ref\rBeBa{O. Babelon and
D. Bernard, Phys. Lett. 260B (1991) 81;
{\it ``Symmetries of the Heisenberg models"},
in preparation.}.

\bigskip
\noindent {\bf 2a) What are the dressing transformations?}

\noindent {\bf (i) Equations of motion and Lax connexions.}
Suppose that the equations of motion of a set of fields
$\phi$ are described by a set of non-linear differential
equations. Suppose moreover that these equations
admit a Lax representation. This means that there exists
a field dependent connexion,
called  the Lax connexion, $\CD_\mu$,
$$\CD_{\mu}=\d_{\mu}-A_{\mu}[\phi],$$
such that the equations of motion are equivalent to the zero curvature
condition for $\CD_\mu$,
\eqn\eii{
\BBL\ \CD_{\mu}\ ,\ \CD_{\nu}\ \BBR\ =\ 0}
The Lax connexion takes value
in some Lie algebra $\CG$ with Lie group $G$.

Notice that,
thanks to the zero-curvature condition, the Lax connexion is a pure
gauge; i.e. there exists a $G$-valued function $\Psi(x,t)$
such that:
\eqn\eiv{
\BL\ \d_{\mu} - A_{\mu}\ \BR\Psi\ =\ 0 \qquad
{\rm or}\qquad
A_{\mu}\ =\ \BL\d_{\mu}\Psi\BR\ \Psi^{-1} }
The function $\Psi(x,t)$ is defined up to a right multiplication by
a space-time independent group element. This freedom is fixed
by imposing a normalization condition on $\Psi$; e.g.
$\Psi(x_0)=1$ for some point $x_0$.

\noindent
{\bf (ii) Construction of the dressing transformations.}
The dressing transformations are non-local gauge
transformations acting on the Lax connexion $A_{\mu}\to A^g_{\mu}$
and leaving its form invariant. They thus induce a transformation
of the field variables $\phi\to\phi^g$ mapping a solution of
the equations of motion into another.

They are constructed as follows.
First let us study the set of gauge transformations mapping the
Lax connexion $A_{\mu}$ on a given connexion $A_{\mu}^g$.
Suppose that there exist two $G$ valued functions,
$\Th^g_+$ and $\Th^g_-$, such that:
\eqn\eix{
A_{\mu}^g\ =\ \(\d_{\mu}\Th^g_{\pm}\)\ \Th^g_\pm\,^{-1} +
\Th^g_{\pm}\ A_{\mu}\ \Th^g_\pm\,^{-1} }
Since $A_{\mu}$ is a pure gauge, $A_{\mu}=\(\d_{\mu}\Psi\) \Psi^{-1}$,
$A_{\mu}^g$ is also a pure gauge, $A^g_{\mu}=
\d_{\mu}\(\Th^g_{\pm}\Psi\) \(\Th^g_{\pm}\Psi\)^{-1}$. This implies that
$\(\Th^g_+\Psi\)$ and $\(\Th^g_-\Psi\)$ differ by a right multiplication
by a space-time independent group element
which we denote by $g$. Equivalently:
\eqn\ex{ \Th^g_-\,^{-1}\ \Th^g_+\ =\ \Psi\ g\ \Psi^{-1} }
The main idea underlaying the dressing tranformations
is to consider eq. \ex\ as a factorization problem;
i.e. we look for two subgroups $B_{\pm}\subset G$ such that any
element $h\in G$ admits a unique decomposition $h=h_-^{-1}\ h_+$
with $h_{\pm}\in B_{\pm}$. The requirement that $\Th_{\pm}$
belongs to $B_{\pm}$ then
specify them uniquely from eq. \ex .
The subgroups $B_{\pm}$ are found by demanding that the transformations
\eix\ preserve the form of the Lax connexion.

The  factorization problem in $G$,
\eqn\exi{ g=g_-^{-1}\ g_+ \quad
{\rm with}\quad g_{\pm}\in B_{\pm} }
is a called an
algebraic Riemann Hilbert problem (by analogy with the classical
Riemann Hilbert problem).
For the dressing transformations
to be well-defined this decomposition as to be unique.

The gauge transformation \eix\ induces a transformation
of the group valued function $\Psi$: $\Psi \to \Psi^g$.
Decompose the group  element $g \in G$
as $g = g_-^{-1}\ g_+$ with $g_{\pm}\in B_{\pm}$,
in the way specified by
the algebraic factorization problem discussed above, then,
\eqn\exiv{
\Psi^g\ =\ \(\Psi g \Psi^{-1}\)_+\ \Psi\ g_+^{-1}\
=\ \(\Psi g \Psi^{-1}\)_-\ \Psi\ g_-^{-1}}
The transformation \exiv\ is well defined on the phase space
because it preserves the normalization condition
$\Psi(x_0)=1$.

\noindent
{\bf (iii) The composition law for the dressing transformations.}
It is not the compostion law in $G$
\rSem\ \ref\rBA{J. Avan and M. Bellon, Phys. Lett. B213 (198) 459}.
Let $g,\ h\in G$ with decomposition,
$g=g_-^{-1}\ g_+$ and $h=h_-^{-1}\ h_+$, their composition law
in the dressing group is:
\eqn\exvii{
(h_+,h_-)\bullet (g_+,g_-)\ =\ (h_+g_+ , h_-g_- )}
In particular, the plus and minus components commute.
We denote by $G_R$ the new group equipped with
this multiplication law. The group law \exvii\ can be derived
by using the dressing of $\Psi$. First, we dress $\Psi\to\Psi^g$
by the $g=g_-^{-1}g_+$ according to eq. \exiv. Then, we dress
$\Psi^g\to\(\Psi^g\)^h$ by $h=h^{-1}_-h_+$:
\eqn\exii{
\(\Psi^g\)^h\ = \Th^h_{\pm}\ \Psi^g\ h^{-1}_{\pm}
\quad {\rm with}\quad
\Th^h_\pm = \(\Psi^g\ h\ \Psi^g\,^{-1}\)_\pm }
Using the definition \exiv\ of $\Psi^g$, the factorization of
$\Psi^g h \Psi^g\,^{-1}$ can be written as follows:
$$
\Th^h_-\,^{-1}\Th^h_+ =
\Psi^g h \Psi^g\,^{-1} =
\Th^g_-\Psi (h_-g_-)^{-1}(h_+g_+) \Psi^{-1} \Th^g_+
$$
This implies that
$(\Th^h_-\Th^g_-)^{-1}(\Th^h_+\Th^g_+) =
\Psi (h_-g_-)^{-1}(h_+g_+) \Psi^{-1}$, or equivalently:
$$\BL\Psi (h_-g_-)^{-1}(h_+g_+) \Psi^{-1}\BR_\pm\ =\
\(\Psi^g h \Psi^g\,^{-1}\)_\pm \(\Psi g \Psi^{-1}\)_\pm \ = \
\Th^h_\pm\ \Th^g_\pm $$
This proves eq. \exvii .

For infinitesimal transformations,
$g\simeq 1+X$, $X\in\CG$, and $g_{\pm}\simeq 1+X_{\pm}$ with $X=X_+-X_-$,
the dressing transformations are:
\eqn\exviiI{
\de_X\ \Psi\ = Y_{\pm}\ \Psi - \Psi X_{\pm}
\qquad {\rm with}\qquad Y_{\pm}= \(\Psi X\Psi^{-1}\)_\pm .}
The composition law is, for $X,Z\in\CG$ :
\eqn\exviii{
\BBL\ X\ ,\ Z\ \BBR_R\ =\
\BBL\ X_+\ ,\ Z_+\ \BBR\ -\
\BBL\ X_-\ ,\ Z_-\ \BBR\ }
This defines a new Lie algebra $\CG_R$ which is the Lie algebra of $G_R$.

\bigskip
\noindent {\bf 2b) Few of their properties.}

\noindent
{\bf (i)} {\it They are non-local.} This is obvious from their definition
as $\Psi$ is non-local. The dressing transformations
can be used to construction solutions of the soliton equations having
non-trivial topological numbers from solutions with trivial
topological numbers:
\eqn\exxx{ \phi(x)\ \to\ \phi^g(x) \quad;\quad \forall g\in G_R ;}
In particular, by dressing local conserved currents,
the dressing transformations provide a way to
construct non-local conserved currents :
\eqn\exxxi{J_{\mu}(x,t)\to J^g_\mu(x,t)\qquad;\quad\forall g\in G_R.}
In the quantum theories, these non-local currents are turned into the
generators
of the quantum group symmetries.

\noindent
{\bf (ii)} {\it They induce Lie Poisson actions.}
The dressing transformations induce an action of
the group $G_R$ on the space of solutions of the classical equations
of motion, i.e. on the phase space.
This action is (in general) compatible with the Poisson
structure; more precisely, it is a Lie Poisson action.
It means that the Poisson brackets transform covariantly
if the group $G_R$ is equipped with a non-trivial Poisson structure.
This Poisson structure, which, by construction, is compatible with
the multiplication in $G_R$, turns
the group $G_R$ into a Lie Poisson group.

Let us be a more precise. Denote by $\CP$ the phase space and
by $\{,\}_\CP$ the Poisson bracket on it.
The dressing transformations define an action of $G_R$
on the function over the phase space: $f^g(x)=f(g^{-1}\cdot x)$
for $f\in {\rm Funct}(\CP)$ and $x\in\CP$.
Suppose that the group $G_R$ of
is equipped with a Poisson bracket which we denote by
$\{,\}_{G_R}$.
The statement that the
dressing transformation are Lie Poisson action is equivalent to
the covariance of the Poisson brackets:
\eqn\elpa{
\{f_1,f_2\}_\CP^g\ =\
\{f^g_1,f^g_2\}_{\CP\times G_R} \qquad \forall\ f_1,f_2 \in {\rm Funct}(\CP)\
;\ \forall\ g\in G_R }
The Poisson bracket on $\CP\times G_R$ is the product Poisson structure.

\noindent
{\bf (iii)} {\it  A standard example.}
As is well known, a Lie group $G$ can be equipped with the
following Poisson bracket (Sklyanin's Poisson bracket)
\ref\Skl{E. Sklyanin, {\it ``On the complete
integrability of the Landau-Lifshitz
equation"}, preprint LOMI E-3-79 (1980) Leningrad.}:
\eqn\extra{
\Bigl\{\ \Psi(x)\ \x,\ \Psi(x)\ \Bigr\}\ =\
\BBL\ r^\ep\ ,\ \Psi(x)\x\Psi(x)\ \BBR }
with matrices
$r^\ep$, $\ep=\pm$, solutions of the classical Yang-Baxter equation.
By $G$-invariance, in eq. \extra\ we can choose any of two
solutions $r^+$ or $r^-$ of the classical Yang-Baxter equation
provided that their difference is the tensor Casimir $C=r^+-r^-$.
A direct computation \rSem\ \ref\rSeBB{O. Babelon, unpublished}
shows that the Sklyanin's Poisson brackets
are covariant under the transformation \exiv , $\Psi\to\Psi^g$,
only if there are non-trivial Poisson brackets among the $g$'s
but vanishing Poisson brackets between the $g$'s and the fields $\Psi$.
The Poisson brackets in $G_R$ are:
\eqna\exx
$$\eqalignno{
\Bigl\{\ g_+\ \x_,\ g_+\ \Bigr\}\ &=\
\BBL\ r^{\pm}\ ,\ g_+\x g_+\ \BBR &\exx a\cr
\Bigl\{\ g_-\ \x_,\ g_-\ \Bigr\}\ &=\
\BBL\ r^{\pm}\ ,\ g_-\x g_-\ \BBR &\exx b\cr
\Bigl\{\ g_-\ \x_,\ g_+\ \Bigr\}\ &=\
\BBL\ r^-\ ,\ g_-\x g_+\ \BBR &\exx c\cr
\Bigl\{\ g_+\ \x_,\ g_-\ \Bigr\}\ &=\
\BBL\ r^+\ ,\ g_+\x g_-\ \BBR &\exx d\cr}$$
For $g=g_-^{-1}\ g_+$, the Poisson brackets are the
Semenov-Tian-Shansky brackets:
\eqn\exxi{\eqalign{
\Bigl\{\ g\ \x_,\ g\ \Bigr\}\ =\
(g\x1)r^+(1\x g) &+ (1\x g)r^-(g\x 1) \cr
 - (g\x g)r^{\pm} &- r^{\mp}(g\x g) \cr}}
It is easy to check that the multiplication in $G_R$
(not in $G$) is a Poisson mapping for the Poisson structure
defined in eq. \exxi , or in eq. \exx{}.
Therefore $G_R$ is a Poisson Lie group and the actions \exiv\
are Lie Poisson actions.

\bigskip
\newsec{AN EXAMPLE OF DRESSING TRANSFORMATIONS: CURRENT ALGEBRAS.}

\def\nj{{J^{(1)}}}
\def\oq{Q_0}
\def\nq{Q_1}

We illustrate the general construction explained in the previous
section on the example of the classical current algebras. We
essentially follow \ref\rUne{K. Uneo and Y. Nakamura,
Phys. Lett. 117B (1982) 208}. Dressing transformations
in the Toda theories were traited in \rBeBa .

\bigskip
\noindent
{\bf 3a) The equations of motion.}

The field variables are one-forms, denoted by $J_{\mu}(x)$,
valued in a semi-simple Lie algebra
$\CG$: $J_\mu(x)= \sum_a\ J_{\mu}^a(x)t^a$ where
$t^a$, $a= 1, \cdots, {\rm dim}\CG$, form a basis of $\CG$
\foot{We suppose the $t^a$ orthonormalized.
We use the convention: $\[t^a,t^b\]=f^{abc}t^c$ where $f^{abc}$
denote the structure constants of $\CG$.}.
By definition the equations
of motion impose to $J_\mu(x)$ to be a curl-free conserved current:
\def\*{\star}
\eqn\ei{\eqalign{
& \d_{\mu}J^a_{\mu}(x) \ = 0 \cr
&\d_{\mu}J_{\nu}^a(x) - \d_{\nu}J^a_{\mu}(x)
+ f^{abc}J^b_{\mu}(x)J^c_{\nu}(x) = 0 \cr}}
The equations of motion \ei~ admit a Lax representation:
they are equivalent to the zero curvature condition,
$\[ \CD_{\mu}(\la), \CD_{\nu}(\la)\]=0$, for the
connexion $\CD_{\mu}(\la)$,
\eqn\eii{
\CD_{\mu}(\la)\ = \ \d_{\mu} + \frac{\la^2}{\la^2-1}\ J_{\mu}(x)
	+ \frac{\la}{\la^2-1}\ \ep_{\mu \nu}J_{\nu}(x) }
The Lax connexion is an element of the loop algebra
$\hat \CG = \CG\x C[\la,\la^{-1}]$.

\noindent {\bf Remark 1:}
The linear problem
$\(\d_\mu - A_\mu(x)\) \Psi(x)=0$
associated to the Lax representation \eii\
is equivalent to the following ($\ep_{\mu\nu}\ep^{\nu\sig}=\de_\mu^\sig$):
\eqn\eprob{
\BL \d_\mu - \la\ep_{\mu\nu}\d_\nu -\la \ep_{\mu\nu}J_\nu\BR\ \Psi(x) = 0}

\noindent {\bf Remark 2:}
In the light-cone components of the Lax connexion are $(\ep_\pm^\pm=\pm1)$:
\eqn\eapm{ A_\pm\ = - \frac{\la}{\la\mp1}\ J_\pm }
The Lax connexion is therefore  completely characterized by
the following two conditions: i) $A_\pm$ have a simple pole at $\la=\pm1$
(we then set ${\rm Res}_{\la=\pm1}\ A_\pm = \mp J_\pm$) and ii)
$A_\pm(\la =0) =0$.
Therefore, for a gauge transformation to be a symmetry it only has
to preserve these two conditions.

\noindent {\bf Remark 3:}
The gauge condition $A_\pm(\la=0)=0$ implies that $\Psi(\la=0)$ is
space-time independent.
In the following we fix the gauge on $\Psi$ by setting $\Psi(\la=0)=1$.
Moreover we have:
\eqn\ecur{
 J_\nu(x) = \d_\la\ \BL \ep_{\nu\mu}\ A_\mu(x) \BR_{\la=0}
  = \d_\la\ \BL \ep_{\nu\mu}\(\d_\mu\Psi\)\Psi^{-1} \BR_{\la=0}.  }

\bigskip
\noindent {\bf 3b) Dressing transformations and
the Riemann Hilbert problem.}

First, because the Lax connexion takes value in the loop algebra,
we have to define the factorization problem \exi\ in the loop group.
It can be formulated as follows:
Let $\Ga$ be a contour around the origin $\la=0$. Denote by
$\Ga_-$ $(\Ga_+)$ the exterior (interior) domain of $\Ga$. We choose
$\Ga$ such that the points $\la=\pm1$ belong to $\Ga_-$.
The factorization problem
consists in factorizing any regular element of
the loop group, $G(\la)$, $\la\in \Ga$,
into the product of two $\la$-dependent group elements $G_\pm(\la)$
respectively analytic on $\Ga_\pm$:
\eqn\eRH{ G(\la)\ =\ G^{-1}_-(\la)\ G_+(\la)
\qquad;\qquad \la\in \Ga }
This is the Riemann-Hilbert factorization problem. It is known that
it admits a unique solution up to a left multiplication,
$G_\pm \to M G_\pm$, by a $\la$-independent group element $M$.
The definition of the Riemann-Hilbert factorization is cooked up such that
the transformations we will now define are symmetries of the equations
of motion of the classical current algebras.

To dress the Lax connexion \eii , we follow the general procedure:\par
\noindent {\bf (i)} We pick up an element $G(\la)$ of the loop group.
We fix the gauge in the Riemann-Hilbert factorization by imposing
$G_+(\la=0)=1$.

\noindent {\bf (ii)}
We define $\Th^G(\la)= \Psi\ G(\la) \Psi^{-1}$ and factorize
it according to the Riemann-Hilbert problem:
\eqn\eci{\Th^G(\la)= \Psi\ G(\la) \Psi^{-1}=
	\Th^G_-(\la)^{-1}\ \Th^G_+(\la) .}
We impose the gauge condition $\Th^G_+(\la=0)=1$.
The solution to eq. \eci\ is then unique.

\noindent {\bf (iii)} We define the dressed Lax connexion by:
\eqn\ecii{
A^G_\mu\ =\ \(\d_\mu \Th^G_\pm\)\ \Th^G_\pm\,^{-1} \ +\
\Th^G_\pm\ A_\mu\ \Th^G_\pm\,^{-1}  .}
Because we can implement the dressing either
using $\Th_+$ or using $\Th_-$, it easy to check that the dressed
connexion $A^G_\mu$ possesses the same poles with the same orders than
the original connexion $A_\mu$. The gauge conditions we choose for the
Riemann-Hilbert factorization also ensure that $A^G_\mu$ satisfy the same
gauge condition as $A_\mu$. Therefore, the dressing transformation
$A_\mu \to A^G_\mu$, preserving the structure of the Lax connexion,
induce a symmetry of the equations of motion. The dressed currents
$J^G_\mu(x)$ are defined via the eq. \ecur\ with $A^G_\mu$ instead
of $A_\mu$:
\eqn\edrj{
J^G_\mu(x)\ =\ J_\mu(x)\ +\ \ep_{\mu\nu}\d_\nu
\BL\d_\la\Th_+^G\BR_{\la=0} }.

\noindent {\bf (iv)}
For infinitesimal transformations, $G(\la)=1+X(\la)+\cdots$,
where $X(\la)\in \hat\CG$, $X(\la)=X_+(\la)-X_-(\la)$ with $X_\pm(\la)$
analytic in $\Ga_\pm$, the dressings are:
\eqn\edii{\eqalign{
\de_X\ A_\mu \ &=\ \d_\mu\ Y_\pm\ +\ \BBL Y_\pm,A_\mu\BBR \cr
\de_X\ \Psi\ &=\ Y_\pm\ \Psi\ -\ \Psi\ X_\pm \cr}}
with $Y(\la)=\(\Psi X(\la) \Psi^{-1}\)(\la)= Y_+(\la)-Y_-(\la)$.
In particular for the current:
\eqn\ediii{
\de_X\ J_\mu\ =\ \ep_{\mu\nu}\d_\nu\BL\d_\la Y_+\BR_{\la=0} }

\bigskip
\noindent{\bf 3c) Non-local conserved currents.}

The problem consists now in solving the Riemann-Hilbert factorization,
eq. \eRH .
The main point is that we will find differential equations
which solve this problem recursively. In the following we restrict
ourselves to the dressing of the current $J_\mu(x)$.

\noindent {\bf (i)} {\it Projection on $\hat \CG_+$}. Recall that by
definition, eq. \eRH, $\hat \CG_+$ is the algebra
of $\CG$-vector fields regular at
the origin $\la=0$. Therefore, if $Y(\la)$ is an element of the loop
algebra $\hat \CG$, its projection $Y_+(\la)$ on $\hat \CG_+$ is:
\eqn\ecx{ Y_+(\la)\ =\ \oint_\Ga\ \frac{dz}{2i\pi}\
	\frac{ Y(z)}{z-\la}\qquad;\qquad \la\in\Ga_+ }

\noindent {\bf (ii)} {\it The dressing transformations act on $J_\mu$
by non-local gauge transformations.} The dressing of $J_\mu$ is defined in
eq. \edrj , or its infinitesimal form \ediii.
To compute it we use the explicit expression of $Y(\la)$ and the projector
\ecx :
\eqn\ecxi{\eqalign{
 \d_\mu\BL\d_\la Y_+\BR_{\la=0} \ &=\
\oint\frac{dz}{2i\pi z^2} \BBL \(\d_\mu\Psi(z)\)\Psi^{-1}(z)\ ,\
	\Psi(z) X(z) \Psi^{-1}(z) \BBR \cr
&=\ \ep_{\mu\nu}\BL\ \d_\nu Z_+\ +\ \[{J_\nu,Z_+}\]\ \BR \cr}}
with:
\eqn\ecxii{ Z_+\ =\ \oint \frac{dz}{2i\pi z} \BL\Psi(z)X(z)\Psi^{-1}(z)\BR.}
To derive the last equation we used the linear problem in the form \eprob .
The variation of $J_\mu$ is therefore:
\eqn\ecxbis{
\de_X\ J_\mu\ =\
\d_\mu Z_+\ +\ \[{J_\mu,Z_+}\] .}

\noindent {\bf (iii)} {\it Recursion relation for $Z_+$.}
The last step consists in solving for $Z_+$ by recursion. Let
$X\in{\hat \CG}$ be $X^n(\la)=v\la^{-n}$ with
$n=0,1,\cdots$ and $v\in\CG$. Denote by $Z^n$ the corresponding
solution to eq. \ecxii :
\eqn\ecxiii{
Z^n\ =\ \oint \frac{dz}{2i\pi z} \BL\Psi(z)v\Psi^{-1}(z)\BR z^{-n}.}
Then using once more the differential equation \eprob, we have:
\eqn\ecxiv{ \d_\mu Z^{n+1}\
=\ \ep_{\mu\nu}\BL\ \d_\nu Z^n\ +\ \[{J_\nu,Z^n}\]\ \BR .}
As advertised, this solves recursively the Riemann-Hilbert
problem. The dressed currents, $\de^n_v J_\mu$,
 are recursively defined by
eqs. \ecxbis\ and \ecxiv . This recursive construction is equivalent to the
construction of ref. \ref\BIZZ{E. Brezin et al, Phys. Lett. 82B (1979) 442}.
The conservation law for the dressed
currents $\de^n_v J_\mu$ can be checked directly.

\noindent {\bf (iv)}
{\it The two first conserved currents.} The first ones are the local
currents $J_{\mu}^a(x)$ since, for $X=v\in\CG$,
\eqn\ecxv{ \de^0_v\ J_\mu(x)\ =\ \BBL J_\mu(x), v\BBR .}
For $X=v\la^{-1}$, $v\in\CG$, we have $\d_\mu Z^1=\ep_{\mu\nu}
\[J_\nu,v\]$, or equivalently,
\eqn\ecxvi{ Z^1(x)=\BBL\Phi(x),v\BBR\quad {\rm with}\quad
\Phi(x)=\int_{\CC_x} \* J }
where $\CC_x$ is a curve ending at the point $x$.
The dressing of $J_\mu$ is:
\eqn\ecxvii{ \de_v^1 J_\mu(x) = \ep_{\mu\nu}\BBL J_\nu(x),v\BBR
+ \BBL J_\mu(x),\BBL\Phi(x),v\BBR\BBR.}
In particular, projecting on the adjoint representation, we find
the following non-local conserved currents:
\eqn\ecxix{
\eqalign{ & f^{abc} \de^1_{t^b}\ J^c_\mu(x) \propto \nj^a_\mu(x) \cr
& \nj^a_{\mu}(x)\ = \ \ep_{\mu \nu}J^a_{\nu}(x)
+ \half f^{abc} \ J^b_{\mu}(x)\ \Phi^c(x) \cr}}

\bigskip
\newsec{QUANTIFICATION: YANGIANS IN MASSIVE CURRENT ALGEBRAS.}

We use the example of the massive current algebras in order to
describe how non-local conserved currents can be defined in
a non-perturbative way and to illustrate few of their
properties, (e.g. how they act on the states, on the
fields, etc...). But the approach is more general, see e.g. ref. \rBL.

The currents $\nj^a_{\mu}(x)$ are the
currents we want to quantize.
There are different ways to specify the quantum theory, e.g.
by defining it on the lattice, or as perturbation of its
U.V. fixed point, etc... Here we use an alternative approach:
we look for the conditions that we have to impose on the
operator algebra in order to be able to define the quantum
non-local conserved currents.
Therefore, we are interested in a quantum models satisfying
the following hypothesis:

\noindent {\bf (a)} There exist quantum local conserved currents,
$J^a_\mu(x)$, taken values in the Lie algebra $\CG$ :
\eqn\edi{ \d_\mu\ J^a_\mu(x)\ =\ 0 .}
Furtheremore, because the currents $J^a_\mu$ have to be one-forms,
we impose that they have
scaling dimensions one.

\noindent {\bf (b)} The currents $J^a_\mu(x)$ satisfy the quantum
version of the equations of motion \ei ; i.e.
the quantum currents are curl-free:
\eqn\eo{
\d_{\mu}J^a_{\nu}(x) - \d_{\nu}J^a_{\mu}(x)
+  \ f^{abc}\ :J^b_{\mu}(x)\ J^c_{\nu}(x):\  = 0 }
where the double dots denote an appropriate regularization of
$f^{abc}J^b_\mu(x)J^c_\nu(x)$, e.g. by a point splitting.
This hypothesis imposes constraints on
the operator product expansion (OPE) of the currents.

\noindent {\bf (c)}
The only fields
taking values in the adjoint representation of $\CG$
and having scaling dimensions zero, one or two are either
$J^a_\mu$ or $\d_\nu J^a_\mu$.
This fixes the OPE
$f^{abc}J^b_\mu(x)J^c_\nu(0)$ up to the order $\CO(|x|^{1-0})$:
\eqn\evi{
f^{abc}\ J^b_{\mu}(x)J^c_{\nu}(0) =
\CC^{\rho}_{\mu \nu}(x)\ J^a_{\rho}(0)
+\CD_{\mu \nu}^{\sig \rho}(x) \BL \d_{\sig}J^a_{\rho}(0)\BR
+\CO(|x|^{1-0}) }

The quantum currents $J^a_{\mu}(x)$ satisfying these three hypthesis
generate what could be called a massive current algebra.

\bigskip
{\bf 4a) OPE's in massive quantum current algebras.}

We now show that these hypothesis ensure that
the currents satisfy the commutation relations of a Kac-Moody
algebra but also that they satisfy the following OPE's
\foot{We used the following space-time conventions
$x^{\nu}\equiv (x^0=t,x^1=x)$; $x^{\pm}=x\pm t$; and
$ds^2=\eta_{\mu \nu}dx^{\mu}dx^{\nu}=dt^2 - dx^2$.}:\par
\eqna\eviii
$$\eqalignno{
J_{\pm}^b(x)J^c_{\pm}(0) &=\ -\frac{k\de^{ab}}{8i\pi}\inv{(x^\pm)^2}
\ -\ \frac{f^{abc}}{2i\pi}\frac{J^c_\pm(0)}{x^{\pm}}
+ \CO(|x|^{-0}) &\eviii a \cr
 ~&~\cr
\half f^{abc}\ \BL J_+^b(x)J^c_-(0) &-
 J_-^b(x)J^c_+(0) \BR &\eviii b \cr
& =  \frac{C_{Adj}}{8i\pi}\log\(M^2 x^+x^-\)\BL\d_+J^a_-(0)-\d_-J^a_+(0)\BR
+\CO(|x|^{1-0}) \cr}$$
$C_{Adj}$ is the Casimir of $\CG$ in the adjoint representation and
$M$ is the mass scale. The product $J^a_\pm(x)J^c_\mp(0)$ is
logarithmically divergent.

We solve for the OPE \evi\ using our hypothesis.
The proof goes in few steps:\par
\noindent {\bf (i)} First, locality,
PT-invariance and Lorentz covariance determine
the general tensor form of
$\CC_{\mu \nu}^{\rho}(x)$ and $\CD_{\mu \nu}^{\sig \rho}(x)$.
Notice also that the conservation law
for $J^a_{\mu}$ allows us to choose $\CD_{\mu \nu}^{\sig \rho}(x)$
to be traceless: $\eta_{\sig \rho}\CD_{\mu \nu}^{\sig \rho}(x) =0$.
Therefore, under the conditions (a) to (c),
the generators  $J^a_{\mu}(x)$ of a massive current algebra
satisfy the following OPE's \Lu\ :\par
\eqn\eiv{
\eqalign{
 f^{abc}  \ J^b_{\mu}(x)J^c_{\nu}(0)\ &=\cr
\BL C_1\ x^2\eta_{\mu \nu}x^{\rho} + C_2\ x^2\(x_{\mu}\de_{\nu}^{\rho}
+x_{\nu}\de_{\mu}^{\rho}\) &+ C_3\ x_{\mu}x_{\nu} x^{\rho} \BR
\( J_{\rho}^a(0) + \half x^{\sig}\d_{\sig}J^a_{\rho}(0)\) \cr
 + \BL D_1\ x^{\rho}\(x_{\mu}\de_{\nu}^{\sig}
- x_{\nu}\de_{\mu}^{\sig}\)
& +  D_2\ x^{\sig}\(x_{\mu}\de_{\nu}^{\rho}
- x_{\nu}\de_{\mu}^{\rho}\)\BR \BL \d_{\sig}J^a_{\rho}(0)\BR \cr
& + \CO\( |x|^{1-0} \) \cr }}
The coefficients $C_i$ and $D_i$ only depend on $x^2$.
Furthermore, the conservation law for the currents implies
the following differential equations for the functions $C_i$
and $D_i$ \Lu\ :
\eqna\eva
$$\eqalignno{
x^2\frac{d}{dx^2} C_2 \ &= \ -\half \(C_1 + 5 C_2 \) &\eva a \cr
x^2\frac{d}{dx^2} \(C_1 + C_2 + C_3 \) \
&= \ - \(C_1 +  C_2 + 2 C_3 \) &\eva b \cr }$$
and
\eqna\evb
$$\eqalignno{
x^2\frac{d}{dx^2} D_1 \ &= \ - D_1 - \frac{x^2}{4}\ C_1&\evb a  \cr
x^2\frac{d}{dx^2} D_2 \ &= \ - D_2 - \frac{x^2}{4}\ C_2&\evb b  \cr
x^2\frac{d}{dx^2} \( D_1 + D_2 \) \ &= \  \frac{x^2}{4}\ C_3&\evb c  \cr}$$

\noindent {\bf (ii)} The differential equations \eva{} and \evb{}
do not specify uniquely the
unknown coefficients $C_i(x^2)$ and $D_i(x^2)$. But we can use the
hypothesis on the scaling dimension of the currents to fixe the leading
behaviour of the functions $C_i(x^2)$:
\eqn\edv{ C_i(x^2)\ =\ \frac{\al_i}{(x^2)^2}\ +\ \CO(|x|^{-3-0})
\qquad;\quad i=1,\ 2,\ 3,}
with $\al_i$ some constants. We assume that there is no leading
logarithmic corrections.

Solving the differential equations \eva{} and \evb{}, we find:
\eqna\evii
$$\eqalignno{
C_1(x^2) &= - \frac{\al}{\(x^2\)^2}
+ \CO(|x|^{-3-0})  &\evii a \cr
C_2(x^2) &=  \frac{\al}{\(x^2\)^2}
+ \CO(|x|^{-3-0})  &\evii b \cr
C_3(x^2) &= - \frac{\ga\al}{(x^2)^2} + \CO(|x|^{-3-0}) &\evii c \cr
D_k(x^2)&=\frac{- \al_k}{4 x^2} \log\(-M_k^2\ x^2\)
 + \CO(|x|^{-1-0})\quad;\quad k=1,2 &\evii d\cr}$$
The constants $M_k$ are  related to the mass scale
and $\ga=2\log(M_2/M_1)$.
The constant $\al$ depends on the
normalization of the currents: we fixe the normalization such that
$\al=- \frac{C_{adj}}{2i\pi}$

\noindent {\bf (iii)} We finally impose the zero curvature condition.
{}From eqs. \eiv\ and \evii{}, we have:
\eqn\edvi{\eqalign{
\ep_{\mu\nu} \BBL f^{abc}J^b_\mu(x)J^c_\nu(0)\ &+\
Z(-x^2)\ \(\d_\mu J^a_\nu(0)-\d_\nu J^a_\mu(0)\)\BBR \cr
= -\frac{\al\ga}{4x^2}\frac{x^\mu}{2}&\(x^\rho\ep^{\mu\sig}
+x^\sig\ep^{\mu\rho}\)\(\d_\sig J^a_\rho(0)+
\d_\rho J^a_\sig(0)\) \cr}}
with $Z(-x^2)=\frac{\al}{4}\log(-M_1M_2x^2)$.

The curl-free equation \eo~ is then an immediat consequence
of \edvi\ if $\ga$ vanishes.
The normal order in \eo~ is defined in such a way to cancel the
logarithmic divergence in $f^{abc}J^b_{\mu}J^c_{\nu}$.
Therefore, the curl-free equation
fixes the two mass scale to be equal
$$ \ga\ =\ 2\log(M_2/M_1)\ =\ 0 $$

The same conclusion could have been reached by imposing the
chiral splitting of the leading terms of the OPE \eiv
\foot{The case $\ga\not= 0$ is also quite interesting; it
probably corresponds to the 2D $O(n)$ models. In particular, in this
case the leading terms of the OPE of the currents do not
satisfy the chiral splitting. In other words the chiral components
of the currents, $J_-^a$ and $J_+^a$, are mixed in the leading
terms of the OPE.}; the approach based on the chiral splitting
assumption was described in ref. \denis .

{\bf (iv)} {\it Commutation relations of the currents.}
Finally, other current OPE's can be deduced using the same techniques.
In particular we have:
\eqn\edvii{\eqalign{
J^a_\mu(x) J^c_\nu(0)\ =\ & - \frac{ k\de^{ab} }{2i\pi}\inv{(x^2)^2}
\(x_\mu x_\nu-\half x^2\eta_{\mu\nu}\) \cr
&- \frac{f^{abc}}{2i\pi}\inv{x^2}
\( x_\mu\de_\nu^\rho+x_\nu\de_\mu^\rho-x^2\eta_{\mu\nu}x^\rho \)
J^a_\rho(0) \ +\ \CO(|x|^{-0}) .\cr}}
These OPE's reduce to eq. \eviii{a}. The products of the
quantum operators are defined by:
\eqn\edviii{
J^a_\mu(x,t) J^b_\nu(y,t)= \lim_{\ep\to0^+}\
J^a_\mu(x,t+i\ep) J^b_\nu(y,t) .}
Therefore, using $\lim_{\ep\to0^+}\frac{i\ep}{x^2+\ep^2}=i\pi\de(x)$,
 the OPE's \edvii\ implies:
\eqn\edix{\eqalign{
&\BBL J^a_t(x)\ ,\ J^b_x(0)\BBR = f^{abc}J^c_x(0)\de(x)
-  \frac{k}{2} \de^{ab} \de'(x)\cr
&\BBL J^a_t(x)\ ,\ J^b_t(0)\BBR = f^{abc}J^c_t(0)\de(x) \cr
&\BBL J^a_x(x)\ ,\ J^b_x(0)\BBR = f^{abc}J^c_t(0)\de(x) \cr}}
They are the commutation relations of a current algebra: the
light cone component $J^a_\pm$ satisfy the commutation relations
of the affine Kac-Moody algebra $\CG^{(1)}$.

\noindent {\bf Remark 1:}
Two hidden consequences of the definition of the massive current
algebras we choose are: i) their ultra-violet limits
are WZW models with $\CG^{(1)}\x\CG^{(1)}$ symmetry;
and ii) they describe perturbations of affine
Kac-Moody algebras by the perturbing fields
$\Phi_{{\rm pert.}}(x) = \sum_a J^a_{\mu}(x)\ J^a_{\mu}(x)$.

\noindent {\bf Remark 2:}
The massive current algebras are characterized by the level $K$ of
the affine Kac-Moody algebras. However the OPE's \eviii{}~ and the
curl-free equation \eo~ are model independent
in the sense that they do not depend
on the level.

\noindent {\bf Remark 3:}
Because the WZW models are the U.V. fixed point of the
massive chiral algebras, they are also $Y(\CG)$ invariant.
Actually, The WZW models are $Y(\CG)\x Y(\CG)$ invariant
(at least classically \ref\RAb{M. Abdalla Phys. Lett.
152B (1985) 215}). They are
also $U_q(\CG)\times U_q(\CG)$ invariant.
The perturbing field $J^a_\mu J^a_\mu$
breaks these symmetries down to the
diagonal $Y(\CG)$ symmetry times a fractional supersymmetry.
It could be interesting to solve
the WZW models from their non-local symmetries. This will
provide a test of the idea we are trying to develop for
the massive integrable models.

\bigskip
\noindent {\bf 4b) The quantum non-local conserved
currents.}

\noindent {\bf (i) Their definition.}
Having proved that the quantum conserved currents satisfy the
quantum form \eo~ of the equations of motion \ei,
it is now easy to defined the quantum conserved currents
$\nj(x,t)$. We define them by a point splitting
regularization $(\de >0)$:
\eqn\eix{
\eqalign{
\nj^a_{\mu}(x,t)\ &=\ \lim_{\de \to 0^+}\ \nj^a_{\mu}(x,t|\de) \cr
\nj^a_{\mu}(x,t|\de) &= \ Z(\de)\ep_{\mu \nu}J^a_{\nu}(x,t)
+ \half f^{abc}\ J^b_{\mu}(x,t) \phi^c(x-\de,t) \cr}}
where $\phi^c(x,t)$, which satisfies $d\phi^c=\* J^c$, is defined by:
$\phi^c(x,t) = \ \int_{\CC_x}\ \* J^c$
The contour of integration $\CC_x$ is a curve from
$-\infty$ to $x$.

The renormalization constant $Z(\de)$ is fixed
by requiring that $\nj_{\mu}^a(x,t)$ are finite and
conserved.
First it is easily seen from eq. \eviii{b}
that $\nj^a_{\mu}(x,t)$ is finite whenever
$Z(\de)=\frac{\al}{2}\log(\de)+{\rm constant}$. The constant is
fixed by demanding the conservation law for $\nj^a_{\mu}$. (The other
subleading terms in $Z(\de)$ are meaningless.) Using eq. \ex~ we
deduce,
\eqn\exii{
\d_{\mu}\nj^a_{\mu}(x,t|\de)=\half\ep_{\mu \nu}
\BBL Z(\de)\(\d_{\mu}J^a_{\nu}-\d_{\nu}J^a_{\mu}\)(x,t)
+ f^{abc}\ J^b_{\mu}(x,t)J^c_{\nu}(x-\de,t)\BBR }
Therefore, from eq. \eviii{b} or \edvi, we learn that
$\d_{\mu}\nj^a_{\mu}(x,t|\de)$ vanishes when $\de \to 0$ if
$Z(\de)=\frac{\al}{2}\log(M\de) + \CO(\de^{1-0})$.

\noindent {\bf (ii) Non-locality: the braiding relations.}
The non-local character of the currents $\nj(x,t)$ is encoded in their
braiding relations, the equal time commutation relations.
The latter are described as follows:
Let $\Phi(y,t)$ be a quantum field local with respect to the currents
$J^a_{\mu}(x,t)$. Then it satisfies the following
equal-time braiding relations \denis :
\eqna\exiii
$$\eqalignno{
\nj^a_{\mu}(x,t)\Phi(y,t)&\ =  \ \Phi(y,t) \nj^a_{\mu}(x,t)
\quad ; \quad ~~~ \quad {\rm for} \ \ x<y &\exiii a \cr
\nj^a_{\mu}(x,t)\Phi(y,t)&\ =  \ \Phi(y,t) \nj^a_{\mu}(x,t)
- \half f^{abc}\ \oq^b\BL\Phi(y,t)\BR J^c_{\mu}(x,t)\cr
&~~~~~~~~~~~~~~~~~~~~~~~~~~~~~~~~~~~
\quad ; \quad {\rm for} \ \ x>y
&\exiii b \cr}$$
where $\oq^b$ are the global charges associated with the local
conserved current $J^b_{\mu}$. They are the same as on the
lattice, eq. \eacv .

The proof of the braiding relations \exiii~ is the same as the
proof of the braiding relations for disorder fields. It only
relies on the way to deform the contour $\CC_x$ entering in
the definition of the currents $\nj(x,t)$
The relative positions of the contours $\CC_x$  depend if $\nj(x,t)$ acts
first or second: if $\nj(x,t)$ acts first (second) the contour is slightly
under (above) the equal-time slice $t={\rm cst}$, we denote them
$\CC_x^-$ $\({\CC_x^+}\)$. (Remember that product of operators are
defined by time ordering.) The relation \exiii{a} follows because,
in this case, there is no topological obstruction for moving the
contour from the configuration $\CC^+_x$  to the configuration
$\CC^-_x$.
In the case of the relation \exiii{b}, these is
an obstruction for moving the contour $\CC_x^+$ onto the contour
$\CC_x^-$. This implies non-trivial braiding relations.
All the non-locality of the currents $\nj^a_{\mu}(x,t)$ is concentrated in the
fields $\phi^c(x,t)$, eq. \ex. For $x>y$ the exchange relation
between $\phi^c(x,t)$ and $\Phi(y,t)$ is:
\eqn\exiv{
\eqalign{
\phi^c(x,t)\Phi(y,t) &= \int_{z \in \CC_x^+} \* J^c(z)\Phi(y,t)\cr
&=\int_{z \in \ga(y)}\* J^c(z)\Phi(y,t)
+ \int_{z \in \CC_x^-}\* J^c(z)\Phi(y,t) \cr
& = \oq^c\BL\Phi(y,t)\BR + \Phi(y,t) \ \phi^c(x,t) \cr}}
The contour $\ga(y)$ is a small contour surrounding the point $y$.
Plugging back eq. \exiv~ into the definition of the non-local current
$\nj^a_{\mu}$ proves the braiding relations \exiii{b}.

\bigskip
\noindent {\bf 4c) The non-local conserved charges and their algebra.}

Given conserved currents the associated  charges are defined by integrating
their dual forms along some curves. The charges depend weakly
on the contours of integration because the dual forms are  closed.
The global conserved charges acting on the states of the
physical Hilbert space are defined by choosing the domain of
integration to be an equal-time slice. Namely for a current
${\CJ}_{\mu}(x,t)$:
\eqn\exv{
Q \ = \ \int_{t=cst}
	dx \ {\CJ}_t(x,t) }
We denote by $Q_0^a$ and $Q_1^a$ the global charges associated to the
currents $J_\mu^a(x)$ and $\nj^a_\mu(x)$.

The (non-local) conserved charges generate
a non-abelian extension of the two-dimensional Lorentz algebra.
In two dimensions the Poincar\'e algebra which is generated
by the momentum operators $P_{\mu}$ and the Lorentz boosts
$L$ is abelian. The momentum operators $P_{\mu}$
are the global charges associated with the conserved
stress-tensor $T_{\mu \nu}(x)$:
$\d_{\mu}T_{\mu \nu}(x)=0$
The Lorentz boost $L$ is the global charge associated with the
conserved boost current:
\eqn\boots{
L_{\mu}(x)\ = \ \half \ep^{\rho \sig}
\BL x_{\rho} T_{\mu \sig}(x) -
 x_{\sig} T_{\mu \rho}(x)\BR }
The (non-local) charges satisfy the following
algebraic relations:
\eqn\rela{\eqalign{
\BBL\ \oq^a\ ,\ \oq^b \ \BBR = \ f^{abc} \oq^c
\quad &;\quad
\BBL\ \oq^a\ ,\ \nq^b \ \BBR = \ f^{abc} \nq^c \cr
\BBL\  L\ ,\ \oq^a\ \BBR =\ 0
\quad &;\quad
\BBL\  L\ ,\ \nq^a\ \BBR =\ - \frac{C_{Adj}}{4i\pi}\ \oq^a \cr}}

The relations \rela\ are part of the defining relations
of the semi-direct product of the Yangians $Y(\CG)$
by the Poincar\'e algebra. Only the Serre relations
are missing. (They are more difficult to prove because they
involve commutation relations between the non-local charges.)
Moreover, as we will soon show, the comultiplications
are those in $Y(\CG)$.

The three first relations are easily proved.
The last relation is more interesting and can be proved
in geometrical way.
It consists in imposing a Lorentz boost $\CR_{2\pi}$ of angle
$\(i2\pi\)$ to the non-local currents $\nj^a_{\mu}(x,t)$.
It is a rotation of $(2\pi)$ in the Euclidian plane.
Because the currents $\nj^a_{\mu}(x,t)$ are non-local this transformation
does not act trivially on them: the string $\CC_x$
winds around the point $x$.
By decomposing this winding contour into the sum of a
contour from $-\infty$ to $x$ plus a small contour
surrounding $x$ we obtain:
\eqn\rot{
\CR_{2\pi}\ \nj^a_{\mu}(x,t)\  \CR^{-1}_{2\pi}\ =\
\nj^a_{\mu}(x,t) -
\half f^{abc}\ \oq^c\BL\ J^b_{\mu}(x,t) \ \BR }
Integrating the time-component of eq. \rot~ over an
equal-time slice gives
\eqn\rotb{
\CR_{2\pi}\ \nq^a \ \CR^{-1}_{2\pi}\ =\
\nq^a - \half \ C_{Adj}\ \oq^a}
in agreement with the
relations \rela\
because $\CR_{2\pi}=\exp(i2\pi L)$.

\bigskip
\noindent {\bf 4d) Action on the asymptotic states and the S-matrices.}
Non-perturbative results on the $S$-matrices can be deduced
by looking at the action of the quantum charges on the asymptotic
states. The constraints on the $S$-matrices we obtain arise by requiring
that they commute with the non-local charges. These commutation
relations imply algebraic equations which are nothing but the
exchange relations for the quantum symmetry algebra (the Yangians
$Y(\CG)$ in the case of massive current algebras).
In general, these algebraic equations implies non-trivial
constraints on the $S$-matrices which are sometimes enough
to determine them.

\noindent {\it Example: the SO(N) Gross-Neveu models.}
The $SO(N)$ Gross-Neveu models are
equivalent to the $SO(N)$ massive current algebras
at level $K=1$.
In the $SO(N)$ Gross-Neveu models the
fundamental asymptotic particles are Majorana
fermions taking values in the vector representation of $SO(N)$.
In the SO(N)  Gross-Neveu models, the $Y(SO(N))$ charges
acting on the asymptotic fermions are given by:
\eqna\exxvi
$$\eqalignno{
\oq^{kl}\ & = \ T^{kl} &\exxvi a\cr
\nq^{kl}\ & = \ - \  \frac{\th \ (N-2)}{i\pi}\  \(T^{kl}\) &\exxvi b\cr
\De \nq^{kl} & =\nq^{kl}\otimes 1 + 1 \otimes \nq^{kl}
- \sum_n\(T^{kn}\otimes T^{nl}-T^{ln}\otimes T^{nk}\) &\exxvi c\cr}
$$
where the  $T^{kl}$'s  form the vector representation
$\sqcup$ of $SO(N)$:
$\(T^{kl}\)^{mn}=\de^{km}\de^{ln}-\de^{lm}\de^{kn}$.
The charges $\oq^a$ and $\nq^a$ defined in eq. \exxvi~
satisfy the algebra \rela{}; on-shell the boost
operator $L$ acts as $\der{\th}$.
They define an irreducible representation of the $SO(N)$-
Yangians in the vector representation of $SO(N)$.
Eq. \exxvi{c}~ is the comultiplication in $Y((SO(N))$.

Denote by $S(\th_{12})$, $\th_{12}=\th_1-\th_2$,
the S-matrix of the two-fermion scattering. $S(\th)$ acts from
$\sqcup \otimes \sqcup$ into itself. As an $SO(N)$ representation
the tensor product $\sqcup \otimes \sqcup$ decomposes into
$\( {\sqcap \atop \sqcup}+
\quad \perp \quad + \bullet\)$. We denote
by $P_-, \, P_+$ and $P_0$ the respective projectors. By
$SO(N)$-invariance, $S(\th)$ decomposes on these projectors:
\eqn\exxviii{
S(\th)\ = \ \sig_+(\th)P_+ + \sig_-(\th)P_- +\sig_0(\th)P_0}
where $\sig_n(\th)$ are scattering amplitudes.
The non-local charges $\nq^{kl}$ are conserved and therefore
they commute with the S-matrix. For the two-fermion scattering,
the $Y(SO(N))$ exchange relations imply
the following algebraic relations bewteen the scattering amplitudes:
\eqn\exxx{
\frac{\sig_-(\th)}{\sig_+(\th)} =
\frac{\th(N-2)+i2\pi}{\th(N-2)-i2\pi}
\quad ; \quad
\frac{\sig_0(\th)}{\sig_-(\th)} =
\frac{\th+i\pi}{\th-i\pi}}
Eq. \exxx~ determine $S(\th)$ up to an overall function which
could be fixed by closing the bootstrap program \rZZ .

\bigskip
\noindent {\bf 4e) Action on the fields and the field multiplets.}

\noindent {\bf (i) The definition of the action}.
The definitions of  charges acting on the states and
on the fields differ by the choice of the contour along which
the conserved current is integrated.
The charges acting on a field $\Phi(y)$ located at a point $y$ are defined by
choosing the contour of integration $\ga(y)$ from $-\infty$ to
$-\infty$ but surrounding the point $y$:
\eqn\exvi{
Q_k^a\BL\Phi(y)\BR \ = \ \int_{z \in \ga(y)} dz_\mu
	 \ep^{\nu \mu}\ {J^{(k)}}^a_\nu(z) \Phi(y) }
Compare with the lattice definition \eabvi .

For the currents $J^a_{\mu}(x)$ and the charges $\oq^a$
deforming the contour $\ga(y)$ proves that
\eqn\exvii{
Q_0^a\BL\Phi(y)\BR \ = \
Q_0^a\ \Phi(y) \ - \
\Phi(y)\ Q_0^a }

When the currents and the field $\Phi(y)$
are not respectively local the situation is more subtle.
The contour $\ga(y)$ can no more be closed and the action of the charges
on the field is no more a pure commutator. For the non-local conserved currents
$\nj(x)$ the relation between the global charges \exv~ acting on
the states and the charges \exvi~ acting on the fields is the
following:
\eqn\exviii{
Q_1^a\BL\Phi(y)\BR \ = \
Q_1^a\ \Phi(y) \ - \
\Phi(y)\ Q_1^a
\ +\ \half f^{abc}\
Q_0^b\BL\Phi(y)\BR\  Q_0^c}
The proof of eq. \exviii~ consists in decomposing the contour
of integration $\ga(y)$ into the difference of two contours
$\ga_+$ and $\ga_-$ which are respectively above
and under the point $y$, and in using the braiding
relation \exiii{b}~ when the current $J^b(x)$ is on $\ga_-$.

\noindent {\bf (ii) The comultiplications.}
We now derive the comutiplication from the braiding relations.
The comultiplications just encode how the charges act
on a product of fields, say $\Phi_1(y_1)\Phi_2(y_2)\cdots$.
We denote them by $\De$.
In the case of the charges $\oq^a$ and for fields
$\Phi_n(y_n)$ which are local with respect to the currents $J_{\mu}^a(x)$
all the contours can be deformed without troubles and we have:
\eqn\exix{\eqalign{
Q_0^a\BL\Phi_1(y_1)\Phi_2(y_2)\BR \ &= \
Q_0^a\BL\Phi_1(y_1)\BR \Phi_2(y_2) \ + \
\Phi_1(y_1) Q_0^a\BL\Phi_2(y_2)\BR \cr
\De \oq^a & =\ \oq^a\otimes 1 + 1\otimes \oq^a \cr}}
It is the standard Lie algebra comultiplication as it should be.

In the case of the non-local charges $\nq^a$ the standard
comultiplication is deformed due to the non-trivial
braiding relations between the non-local currents and the fields.
Let $\Phi_n(y_n)$ be quantum fields local with respect to the
currents $J^a_{\mu}(x)$. Then we have the following
comultiplication for the non-local conserved charges $\nq^a$:
\eqna\exx
$$\eqalignno{
Q_1^a\BL\Phi_1(y_1)\Phi_2(y_2)\BR \ = \ &
Q_1^a\BL\Phi_1(y_1)\BR\ \Phi_2(y_2) \ + \
\Phi_1(y_1)\ Q_1^a\BL\Phi_2(y_2)\BR \cr
& \ -\ \half f^{abc}\
Q_0^b\BL\Phi_1(y_1)\BR Q_0^c\BL\Phi_2(y_2)\BR &\exx a\cr
\De Q_1^a \ = \ &
Q_1^a\otimes 1 \ + \
1 \otimes Q_1^a
\ -\ \half f^{abc}\
Q_0^b\otimes Q_0^c &\exx b\cr}$$
Eqs. \exix\ and \exx{}\ are the comultiplication in $Y(\CG)$.
Equation \exx{a}~ can be proved by decomposing the contour
$\ga_{12}$ used in defining the action of $\nq^a$ on the product
$\Phi_1(y_1)\Phi_2(y_2)$. The contour $\ga_{12}$ is surrounding the two
points $y_1$ and $y_2$. It decomposes into the sum of two contours
$\ga_1$ and $\ga_2$ surrounding $y_1$ and $y_2$ respectively.
But on the contour $\ga_2$ we have to use the braiding relations \exiii~
in order to pass the string $\CC_z$ through the point $y_1$.
Eq. \exx~ can also be proved starting from the
graded commutators \exviii.

\noindent {\bf (iii) The field multiplets.}
To any (local) field  $\Phi(x,t)$ is associated a multiplet which
is constructed by acting on the field with as many charges
as possible:
\eqn\edx{ Q^{A_1}\ \cdots\ Q^{A_P}\ \Phi(x,t) }
with, in the case $Y(\CG)$ symmetry, $Q^A=Q^a_0$, $Q_1^a$ or any element
of the algebra generated by them. By construction, the fields \edx\
form a field multiplet in the sense of eq. \eabi .
In general the field multiplets are infinite
dimensional.

The main property of the field multiplets resides,
(assuming the knowledge of the action on the asymptotic states),
in the fact that if the field $\Phi(x,t)$ is known, then all its descendents,
$Q^{A_1}\ \cdots\ Q^{A_P}\ \Phi(x,t) $ are also known.
In other words, the descendents are completely determined by the data
of the fields $\Phi(x,t)$ and of the values of the charges on the
asymptotic states.

This property follows from the Ward identities
expressing the quantum invariance:
\eqn\edxi{
\De^{(M)}\(Q^A\)\ \vev{ \Phi_1(x_1)\cdots \Phi_M(x_M) } =0 }
where $\De^{(M)}$ the $M^{th}$ comultiplication with
$\De Q^A=Q^A\x1+\Th^A_B\x Q^B$.
Here we have assumed that the vaccuum is quantum group
invariant: $Q^A \ket{0}=0$, $\bra{0} Q^A=0$.
The identity \edxi\ can be formulated on the form factors
The form factors are the matrix elements
of the fields between asymptotic states. By crossing symmetry, only
matrix elements between the vacuum and the asymptotic particles
are relevent. Let us denote by $Z^\al(\th)$ the asymptotic
particles with rapidity $\th$; they form a representation $W$ of
the quantum symmetry algebra.
The form factors of the fields $\Phi_i(x,t)$ are defined by:
\eqn\edxii{
\CF_i^{\al_1,\cdots\al_M}\(\th_1,\cdots,\th_M\)\ =\
\bra{0}\Phi_i^\La(0)\ket{Z^{\al_1}(\th_1)\cdots Z^{\al_M}(\th_M)} .}
On the form factors, the Ward identities \edxi~ become:
\eqn\edxiii{\eqalign{
\bra{0} &\BL Q^A\Phi_i(x)\BR
\ket{ Z^{\al_1}(\th_1),\cdots,Z^{\al_M}(\th_M) }\cr
& =\ - \bra{0} \BL\Th^A_B\Phi_i(x)\BR
\BL \De^{(M)}Q^B \ket{ Z^{\al_1}(\th_1)\cdots Z^{\al_M}(\th_M) }\BR \cr
& = \bra{0}\Phi^\La(x)
\BL \De^{(M)}s(Q^A) \ket{ Z^{\al_1}(\th_1)\cdots Z^{\al_M}(\th_M) }\BR \cr}}
with $s$ the antipode. Eqs. \edxiii\
give the form factors of the field $Q^A(\Phi_i(x,t))$ in terms
of those of the fields $\Th^A_B\Phi_i(x,t)$ and
of the action of the charges $Q^A$ on the asymptotic particles
$Z^\al(\th)$.

\noindent {\it Example: action on the stress-tensor.} In massive
current algebras, the stress-tensor and the
current are in the same $Y(\CG)$ - multiplets. We have:
\eqn\edxv{
Q^a_1\(T_{\mu\nu}\)\ \propto\
\ep_{\mu\rho}\d_\rho J_\nu^a +
\ep_{\nu\rho}\d_\rho J_\mu^a .}
This relation was proved in \ref\rLS{A. Leclair and F. Smirnov,
{\it Infinite quantum symmetries of fields in massive quantum
field theories}, to appear in J. Mod. Phys. A}
using form factor technique, it can aslo
be deduced from the hypothesis we made for defining the massive
current algebras.

\noindent {\bf Remark 1:} The Ward identity \edxiii\ can written for any
element in the enveloping algebra. Choosing a particular element
associated to the square of the antipode leading to the so-called
deformed KZ equations for the form factors \ref\rS{I. Frenkel
and N. Reshetikhin, to appear; F. Smirnov, preprint RIMS-772 (1991)}.

\noindent {\bf Remark 2:} Assuming, as in conformal field
theory,  that the (complete) symmetry algebra possess
free field vertex representations,
the form factors will also admit free field
representations. The Zamoldchikov creation operators $Z^\al(\th)$
as well as the field operators $\Phi^\La(x)$ will be represented
as quantum vertex operators in analogy with the vertex operator
representations of the quantum affine algebras.
This is suggested by the explicit
formula for the form factors found by Smirnov \ref\rSmir{F. Smirnov,
{\it Form factors in completly integrable models of quantum
field theory}, to be published in World Scientific }.
Their generic forms are as follows:
\eqn\effi{\eqalign{
\CF(\th_1,\cdots,\th_M)&=\int\prod_k d\mu(\al_k)\
P(\al_1,\cdots,\al_k\vert\th_1,\cdots,\th_M)\cr
&\ \times \prod_{k<l}G_1(\al_k-\al_l)
\prod_{i<j}G_2(\th_i-\th_j)
\prod_{k,j}G_3(\al_k-\th_j) \cr}}
with $d\mu(\al)$ some integration measure, the functions $G_n$
are some models dependent kernels and $P(\al_k|\th_i)$
are polynomials. Formula \effi\
suggest the following vertex operator representations:
\eqn\effii{
\CF(\th_1,\cdots,\th_M)=\int\prod_k d\mu(\al_k)
\vev{ \CO \prod_k V(\al_k) \prod_i Z(\th_i) } }
where the $V(\al_k)$ 's are ``screening" operators,
the $Z(\th)$ 's are vertex operator representations of the
Zamolodchikov operators and $\CO$ an operator
representing the fields.
The kernel between these operators
can be deduced from the formula \effi .

\noindent {\bf Remark 3:} The braiding relations between the
quantum field multiplets are determined by the quantum symmetries:
the braiding matrices intertwine the quantum symmetries, cf e.g.
eqs. \eabvii\ and \eabviii .
Moreover the braiding relations are scale invariant; i.e.
they are renormalization group invariant.
This is obvious from their definitions but this also follows from the
topological origin of the braiding relations.
The braiding relations just reflect the monodromy of the
field multiplet correlation functions. Therefore, the renormalization
group induces isomonodromy deformations \ref\rIso{See e.g. the review
by M. Jimbo, Proc. of Symposia in Pure Math. 49 (1989) 379}.
The connection between isomonodromy deformations and quantum group
symmetries could provide another starting point for determining
the correlation functions in massive two dimensional quantum
field theories.

\bigskip
\newsec{CONCLUSIONS.}

\noindent {\bf Few open problems:}
Quantum symmetries have been used with some success to
study integrable perturbations of conformal field theories.
Some examples are
\ref\rPert{A. Zamolodchikov, Adv. Studies Pure math. 19 (1989) 641;
N. Reshetikhin and F. Smirnov, Comm. Math. Phys. 131 (1990) 157;
D. Bernard and A. Leclair, Nucl. Phys. B340 (1990) 721;
F. Smirnov, Int. J. Mod. Phys. A6 (1991) 1253;
D. Bernard and A. Leclair, Phys. Lett. B247 (1990) 309;
C. Ahn, D. Bernard and A. Leclair, Nucl. Phys. B 346 (1990) 409;
H. de Vega and V. Fateev, preprint LPTHE-90-36;
V. Fateev, Int. J. Mod. Phys. A6 (1991) 2109; etc...}:
the $\Phi_{(1,3)}$ and the
$\Phi_{(1,2)}$, $\Phi_{(2,1)}$ perturbations of the minimal
conformal models, the $G_K\otimes G_L / G_{K+L}$ cosets models,
the $Z_N$ parafermionic models, and the fractional supersymmetric
models, etc... Most of the results obtained in these papers concern
the S-matrices of these massive models. The main  open problem
is the derivation of the off-shell properties
of the models, (the correlation functions and the form factors),
from their quantum symmetries.
As we mentioned in the introduction, this requires checking
if the quantum symmetries form a complete symmetry algebra or not.
Other few technical problems have been formulated in the previous
sections, most of them as remarks. In particular, the connection
between isomonodromy deformations and quantum group symmetries
could open a new way of solving for the correlation functions.

\bigskip
\noindent {\bf 3D generalizations?}
Let us discuss how these constructions could possibly be generalized
to three dimensions. In two dimensions, quantum group symmetries
require non-local currents: the non-locality, which is reflected
in the equal-time commutation relations, imply
the non-trivial comultiplications. The 2D non-local currents are fields
localized on points but with a ``string" attached to them. The
currents are generically express as products of disorder fields
(the ``wavy string" in the lattice description) by spin fields
(which are local fields). This is analogue to the definition
of the 2D parafermions.

In any dimensions, to have more general symmetry than supersymmetry
we need fields with non-trivial equal-time commutation relations.
In three dimensions, this requires to consider fields localized
on curves (with a sheet attached to them). Once again, examples are
provided by disorder and parafermionic fields. The latters can be
described as follows: consider a group G invariant spin lattice
model in three dimensions. The disorder fields $\mu_g(C)$,
$g\in G$, are defined by splitting all the spin variables
$\sigma$ which leave on a surface $\Sigma_C$ bounded by $C$:
$\sig \to g\sig$. By G-invariance, $\mu_g(C)$ depend weakly on
$\Sigma_C$. The 3D parafermions $\Psi_g(C;x)$ are defined as product of
disorder
fields $\mu_g(C)$ by spin fields $\sig(x)$:
\eqn\ecoi{ \Psi_g(C;x)\ =\ \mu_g(C)\ \sig(x)
\qquad ; \qquad x\in C .}
They satisfy non-trivial commutation relations analogous
to the two-dimensional case.
Anions are particular examples of this construction, with the curve
C a small two-dimensional cone extending to the spacial infinity
\ref\rAnions{F. Wilczek, Phys. Rev. Lett. 48 (1982) 1144;
Y.S. Wu, Phys. Rev. Lett. 52 (1984) 2103;
J. Fr\"olich and P. Marchetti, Comm. Math. Phys. 121 (1989) 121}.
Generalizing eq. \ecoi\ by considering product
of spin fields all along the curve $C$ gives the parafermionic
string which has been considered in the 3D Ising model
\ref\rPD{Vl. Dotsenko and A. Polyakov, Adv. Studies Pure Math.
16 (1988) 171}.

Thus, if quantum group symmetry exists in three dimensions,
it is a theory of quantum fields localized on curves, i.e. a theory
of quantum loops. A (formal) example is given by Polyakov's
string representation of gauge theories in three dimensions
\ref\rPo{A. Polyakov, Phys. Lett. 82B (1979) 247}.
Let $W(C)$ be the Wilson loops:
\eqn\ecoiii{ W(C)\ =\ P\exp\({\oint_C\ A}\) }
where $A$ is the Yang-Mills connection. Define the functional
current $\CJ_\mu(C;x)$ by:
\eqn\ecoiv{ \CJ_\mu(C;x)\ =\ W(C)^{-1}\ \frac{\de}{\de x_\mu}
	W(C) }
This functional current is conserved and curl-free:
\eqn\ecov{\eqalign{
& \frac{\de}{\de x_\mu} \CJ_\mu(C;x) =0\cr
& \frac{\de}{\de x_\mu} \CJ_\nu(C;x') -
 \frac{\de}{\de x'_\nu} \CJ_\mu(C;x) +
\BBL \CJ_\mu(C;x)\ ,\ \CJ_\nu(C;x')\BBR =0\cr
& t_\mu \CJ_\mu(C;x) = 0\cr}}
with $t_\mu$ the vector tangent to the curve $C$ at the point $x$.
Formally the following non-local current, $\CJ^{(1)}_\mu(C;x)$,
localized on the curve $C$ is functionally conserved:
\eqn\ecovi{
\CJ^{(1)}_\mu(C;x)\ =\ \ep_{\mu\nu\rho} t_\nu \CJ_\rho(C;x)
+\half \BBL\ \CJ_\mu(C;x)\ ,\ \CP(C;x)\ \BBR}
with $\de\CP(C;x)/\de x_\mu = \ep_{\mu\nu\rho}t_\nu\CJ_\rho(C;x)$.
The analogy with the 2D current algebras is appealing, cf. eq. \ecxix .
It seems to indicate the possibility of having
generalized non-local symmetry in 3D gauge theories. Unfortunately,
the equations of motion \ecov\ are not very rigorous;
only the discretized lattice version has been proved and,
up to our knowledge, no concrete result on the quantum
continuous case has never been proved.
However, to construct generalized quantum group
symmetry in three dimensions remains a very attractive
challenge.

\bigskip \bigskip
\noindent {\bf Acknowledgements}: It is a pleasure to thank my
collaborators, Olivier Babelon, Giovanni Felder
and Andr\'e Leclair. I thank the organizers of the 91 Cargese school
gor giving me the opportunity to present this lecture in a
very pleasant atmosphere.

\listrefs
\end